\documentclass[longauth]{aaEC}

\usepackage{graphicx}
\usepackage{natbib}
\usepackage{scalerel}

\usepackage[table]{xcolor}

\usepackage{txfonts}
\usepackage[pdfencoding=auto,psdextra]{hyperref}
\hypersetup{
    colorlinks=true,
    linkcolor=blue,
    filecolor=magenta,      
    urlcolor=blue,
    citecolor=blue
}
\makeatletter
\renewcommand*\aa@pageof{, page \thepage{} of \pageref*{LastPage}}
\makeatother

\usepackage[utf8]{inputenc}

\usepackage[switch, modulo]{lineno}
\usepackage{euclid}

\usepackage{enumitem}

\newcommand{\orcid}[1]{} %% define as link to https://orcid.org/#1 if needed

\renewcommand{\degr}{\ensuremath{^\circ}\xspace}
\renewcommand{\arcsec}{\ensuremath{^{\prime\prime}}\xspace}
\newcommand{\euc}{\textit{Euclid}\xspace}

\newcommand{\elvis}{\texttt{ELViS}\xspace}
\newcommand{\sext}{\texttt{SExtractor}\xspace}

\newcommand{\ssofinder}{\texttt{SSOFinder}\xspace}
\newcommand{\streakdet}{\texttt{StreakDet}\xspace}

\begin{document}
%
% Put the title of your paper here:
%
   \title{\Euclid: Detecting Solar System objects in \Euclid images and classifying them  using Kohonen self-organising maps\thanks{This paper is published on behalf of the Euclid Consortium.}}

%% please do not edit the author list -- contact ECEB Bureau for changes
% \newcommand{\orcid}[1]{} %% define as link to https://orcid.org/#1 if needed
% \author{\normalsize F.~Author$^{1}$\thanks{Corresponding author, \email{author@institute.mail}}, S.~Writer$^{2,3}$, A.~Third$^{3}$, F.~Plotter$^{4}$}

% \institute{$^{1}$ Dark Energy Institute, Metropolis, 80th Backyard
%   St., Freedonia\\
% $^{2}$ Center for Large-Scale Structure, University of Somewhere,
% Trembley 48, GoK 23456, Backland\\
% $^{3}$ Institute for Fundamental Research, Rainbow University, 812
% Clear Sky Road, 83456 Rainbow, Wonderland  \\
% $^{4}$ Institute for Astrophysics and Cosmology, UC Gilroy, Major
% Garlic St., Gilroy, CA 93712, USA\\}
%%%% please do not edit the author list -- contact ECEB Bureau for changes
%\newcommand{\orcid}[1]{} %% if already defined in aa.cls: comment, or use renewcommand		   
%%%% please do not edit the author list -- contact ECEB Bureau for changes
%\newcommand{\orcid}[1]{} %% if already defined in aa.cls: comment, or use renewcommand			   
\author{A.~A.~Nucita\orcid{0000-0002-7926-3481}\thanks{\email{nucita@le.infn.it}}\inst{1,2,3}
\and L.~Conversi\orcid{0000-0002-6710-8476}\inst{4,5}
\and A.~Verdier\inst{6}
\and A.~Franco\orcid{0000-0002-4761-366X}\inst{2,1,3}
\and S.~Sacquegna\orcid{0000-0002-8433-6630}\inst{1,2,3}
\and M.~P\"ontinen\orcid{0000-0001-5442-2530}\inst{7}
\and B.~Altieri\orcid{0000-0003-3936-0284}\inst{5}
\and B.~Carry\orcid{0000-0001-5242-3089}\inst{8}
\and F.~De~Paolis\orcid{0000-0001-6460-7563}\inst{1,2,3}
\and F.~Strafella\orcid{0000-0002-8757-9371}\inst{1,3,2}
% ...
}
										   
%%%% please do not edit the affiliation list -- contact ECEB Bureau for changes
\institute{Department of Mathematics and Physics E. De Giorgi, University of Salento, Via per Arnesano, CP-I93, 73100, Lecce, Italy%\label{aff1}
\and
INFN, Sezione di Lecce, Via per Arnesano, CP-193, 73100, Lecce, Italy%\label{aff2}
\and
INAF-Sezione di Lecce, c/o Dipartimento Matematica e Fisica, Via per Arnesano, 73100, Lecce, Italy%\label{aff3}
\and
European Space Agency/ESRIN, Largo Galileo Galilei 1, 00044 Frascati, Roma, Italy%\label{aff4}
\and
ESAC/ESA, Camino Bajo del Castillo, s/n., Urb. Villafranca del Castillo, 28692 Villanueva de la Ca\~nada, Madrid, Spain%\label{aff5}
\and
Institute of Physics, Laboratory of Astrophysics, Ecole Polytechnique F\'ed\'erale de Lausanne (EPFL), Observatoire de Sauverny, 1290 Versoix, Switzerland%\label{aff6}
\and
Department of Physics, P.O. Box 64, 00014 University of Helsinki, Finland%\label{aff7}
\and
Universit\'e C\^{o}te d'Azur, Observatoire de la C\^{o}te d'Azur, CNRS, Laboratoire Lagrange, Bd de l'Observatoire, CS 34229, 06304 Nice cedex 4, France%\label{aff8}
% ...
}     

 \date{Accepted XXX. Received YYY; in original form ZZZ}

% 
% Put your abstract here
%
 \abstract{
  The ESA \Euclid mission will survey more than 14\,000 deg$^2$ of the sky in visible and near-infrared wavelengths, mapping the extra-galactic sky to constrain our cosmological model of the Universe. Although the survey focusses on regions further than 15\degr from the ecliptic, it should allow for the detection of more than about $10^5$ Solar System objects (SSOs). After simulating the expected signal from SSOs in \Euclid images acquired with the visible camera (VIS), we describe an automated pipeline developed to detect moving objects with an apparent velocity in the range of $0.1$--$10\arcsec\,{\rm h}^{-1}$, typically corresponding to sources in the outer Solar System (from Centaurs to Kuiper-belt objects). In particular, the proposed detection scheme is based on \sext software and on applying a new algorithm capable of associating moving objects amongst different catalogues. After applying a suite of filters to improve the detection quality, we study the expected purity and completeness of the SSO detections. We also show how a Kohonen self-organising neural network can be successfully trained (in an unsupervised fashion) to classify stars, galaxies, and SSOs. By implementing an early-stopping method in the training scheme, we show that the network can be used in a predictive way, allowing one to assign the probability of each detected object being a member of each considered class.}
%
% Provide up to five key words:
%
%
% Add short versions of title and author list for page headings
%
   \titlerunning{\Euclid: Detecting and classifying Solar System Objects in \Euclid images}
   \authorrunning{A.~A.~Nucita et al.}
   
   \maketitle
%
%-------------------------------------------------------------------
%
%
%   Start the main text of your paper here
%

\section{\label{sc:Intro}Introduction}
  \Euclid is a space mission of the European Space Agency (ESA) {devoted to the study of the amount and distribution of dark energy and dark matter in the Universe using two cosmological probes \citep{euclid1,2024arXiv240513491E}:} weak gravitational lensing and baryonic acoustic oscillations. \Euclid is located at the second Sun-Earth Lagrange point (L2) and equipped with a 1.2\,m Korsch telescope and two instruments: a visible imaging camera with a pixel scale of \ang{;;0.1} \citep[VIS;][]{cropper2024}, and a near-infrared spectrometer and photometer with a pixel scale of \ang{;;0.3} \citep[NISP;][]{prieto2012, euclid3, macias2016,Jahnke2024}. Both instruments have a field of view of $0.53$\,deg$^2$.

  The satellite carries out an imaging and spectroscopic survey, avoiding Galactic latitudes lower than 30$\degr$ and ecliptic latitudes below $15\degr$, performing a total of 35\,000 pointings. According to the mission requirements, \Euclid imaging detection limits are set to $m_\textrm{AB} = 24.5$ in the single broadband filter (550--900\,nm) of VIS \citep[10\,$\sigma$ detection for a \ang{;;0.3} extended source,][]{cropper2024} and $m_\textrm{AB} = 24$ in the  \YE (900--1192\,nm),  \JE (1192--1544\,nm), and  \HE (1544--2000\,nm) broadband filters of NISP \citep[5\,$\sigma$ detection on a point-like source,][]{schirmer2022,euclid1}. An observing sequence consists of a 565\,s VIS exposure and 111\,s exposures for each of the  \YE, \JE, and \HE filters. The sequence is repeated four times for the same field of view before moving to another set of co-ordinates (see e.g. \citealt{scaramella2022}).

  \Euclid can also study SSOs characterised by a high orbital inclination \citep{carruba, novakovic, terai, chen, namouni}. These objects represent a very interesting population for two main reasons. First, such targets often escape current large surveys, usually covering\footnote{\url{https://minorplanetcenter.net/iau/SkyCoverage.html}} a declination (Dec) range from $-30\deg$ to $+60\deg$. Second, their inclination is apparently inconsistent with the unanimously accepted origin of all bodies in the Solar System having started from a very flattened protoplanetary disc that surrounded the proto-Sun. In particular, as far as the present asteroid belt is concerned, both the average eccentricity ($e$) and the inclination ($i$) of the bodies are too high to be accounted for by planetary perturbations in the present configuration as well as by gravitational scatterings between asteroids \citep{nagasawa}. Although some alternative hypotheses have been put forward to explain these characteristics of the asteroid population (see e.g. \citealt{nagasawa}), a very interesting explanation is provided by the grand tack model \citep{walsh}. According to this model, Jupiter and Saturn, immediately after their formation, would have undergone a double orbital migration (first inwards and then outwards), which would have distorted the original distribution of the orbital parameters ({\it a}, {\it e}, and {\it i}) of the numerous planetesimals with which these two planets interacted. It should be noted that the grand tack coupled with the Nice model \citep{tsiganis} explains several characteristics of the asteroid belt and also of the trans-Neptunian objects \citep{deienno, shannon}, including the presence of a far from negligible number of SSOs with a high orbital inclination.
  
  \Euclid data represents a natural complement to ESA Gaia \citep{gaia2018,spoto2018} and the Vera C. Rubin Observatory Legacy Survey of Space and Time \citep[LSST,][]{lsst}. First, the NIR measurements provided by \Euclid can allow one to characterise the SSO chemical composition more accurately than with Gaia data alone \citep{demeo2009, carry2018}. Second, being located at the Sun-Earth L2 point, the astrometry reported by \Euclid presents a significant parallax compared to contemporaneous ground-based observations (0.01\,AU), which should result in tighter constraints on the orbits of newly discovered SSOs \citep{2007Icar..192..475G,eggl2011}. Simultaneous observations of \Euclid's fields from the ground, by for instance LSST, would provide distance estimates for the observed objects, further constraining their orbits \citep[see][]{eggl2011, euclid4, Rhodes2017}. Third, the hour-long sampling of the rotation light curve of SSOs provided by \Euclid will complement the sparse photometry of Gaia and LSST in studies of the 3-D shape and multiplicity of the objects \citep{durech2015, carry2018}.

  With the current survey design, \Euclid is expected to observe approximately $1.5 \times 10^5$ SSOs. Although most of them are located in the main asteroid belt, between Mars and Jupiter, several thousand Kuiper belt objects should also be detected \citep{carry2018}. Depending on their distance from \Euclid at the time of observations, SSOs will present vastly different apparent velocities \citep[Table~2 in][]{carry2018}. The slowest SSOs appear as point-like sources in the images, whereas the faster ones appear as streaks of various lengths.
  
  Besides its significance for planetary science as a legacy science, identifying and removing asteroids in VIS and NISP images is important for weak gravitational lensing, the core science of \Euclid, by preventing contamination of the shear signal \citep{hildebrandt2017}. Owing to the tremendous amount of data that \Euclid will produce \citep[see e.g.][]{euclid1} and the unique aspects of SSOs compared to the stationary sources at the core of \Euclid data processing, there is a need for dedicated tools to detect and identify SSOs. A first step in this direction was carried out by \citet{lieu2019}, who trained deep convolutional neural networks (CNNs) on simulated VIS images to classify SSOs based on morphological properties and showed the capability of a CNN to separate SSOs from other astronomical sources with an efficiency of $96\%$ down to magnitude 26 and for apparent velocities larger than $10\arcsec\,{\rm h}^{-1}$.

  The detection of fast-moving SSOs (with a typical speed larger than $10\arcsec\,{\rm h}^{-1}$) is handled by a dedicated algorithm \citep[see \streakdet described in][]{streakdet,mikko2}. The software was originally developed to detect streaks caused by space debris in images acquired by an Earth-orbiting facility but also performed well at detecting SSOs in synthetic \Euclid images when combined with a post-processing algorithm to link detected streaks between exposures. \citet{2023A&A...679A.135P} present an improvement in the capability to detect asteroid streaks in \Euclid images by using deep learning.

  In this paper, we describe the design and behaviour of a pipeline, \texttt{SSO-PIPE}, currently developed and maintained at the \Euclid Science Operation Centre in ESAC/ESA but not part of the \Euclid processing function. \texttt{SSO-PIPE} is dedicated to the detection of slow-moving SSOs, with typical speeds lower than $10\arcsec\,{\rm h}^{-1}$, in \Euclid images. The pipeline is based on catalogue registration obtained from VIS exposures and returns output catalogues of candidate SSOs. In principle, candidate SSO co-ordinates can be used in turn as priors for NISP observations from which magnitudes in the {$Y_{\rm E}$}, {$J_{\rm E}$}, and {$H_{\rm E}$} bands could be extracted, thus allowing one to determine the taxonomic class of the object, as described by \citet{popescu}.
  
  Once a moving object is identified in a suite of VIS images, and the purity and completeness of the detection algorithm are assessed, we describe an algorithm designed to classify sources (stars, galaxies, and SSOs) based on their un-parameterised images. The method uses a form of neural network; namely, a self-organising map (SOM) first implemented by \cite{Kohonen,kohonenbook}. The main characteristics of SOM are the simplicity of implementing the algorithm, and the capability to use a classification scheme in an unsupervised fashion. While the usefulness of the former property is self-evident, the latter characteristic implies the advantage that, in contrast to most supervised learning schemes, an SOM may detect unknown features and group data accordingly. 

  The article is organised as follows. In Sect.~\ref{sec:simu}, we describe how we simulate the signal of SSOs in VIS images. The detection algorithm is detailed in Sect.~\ref{sec:pipe}, and its completeness and purity are estimated in Sect.~\ref{sec:disc}. In Sect.~\ref{section:som}, we give details on the SOM algorithm and the data set used to train the network. We then discuss the SOM usage for the SSO classification in Sect.~\ref{sec:somend}. In Sect.~\ref{sec:conclusions}, we address some conclusions.

\section{Simulating Solar System objects in \Euclid images}
\label{sec:simu}

  The VIS instrument on board \euc~is characterised by 36 CCDs arranged in a 6 $\times$ 6 square array, each with 4132 $\times$ 4096 (12\,$\mu$m square) pixels with scale ($p_{\rm fov}$) of \ang{;;0.1}, acquiring images in a single wide band \citep[covering the Sloan filter $r+i+z$ band,][]{cropper2024}. The Euclid Wide Survey is conducted in a step-and-stare tiling mode, in which each 0.57 deg$^2$ field is observed at only one epoch. In the nominal science observation sequence of $4362$\,s, VIS will acquire a series of four 565\,s images of the field, only differing by the optimised dither pattern described in \citet{euclid5} and \citet{scaramella2022}. As a result of the planned mission strategy, any potential SSO will appear as a suite of trails of illuminated pixels convolved with the telescope's point spread function (PSF) in the series of four VIS images. 
  
  We first used the simulations of the sky carried out within the Euclid Consortium, which provide catalogues of stars and galaxies expected to be observed in a given direction up to a certain limiting magnitude. These simulated catalogues come with auxiliary files containing all the necessary information (such as spectral databases for different stellar classes and shapes for any simulated galaxies) required to perform the image construction. The simulation was performed by using \elvis (a \texttt{Python} code developed within the Euclid Consortium, \citealt{elvis}). \elvis constructs the full focal plane image using the aforementioned star and galaxy catalogues.

  For our purposes, it was necessary to provide new functionalities for \elvis devoted to the simulation of SSO trails. All the relevant quantities describing a SSO (such as magnitude, speed, and direction of motion) are extracted from uniform distributions between given limits. In particular, we decided to select random velocities in the range of $0\arcsecf1$-$10\arcsec\,{\rm h}^{-1}$ and an orientation angle, $\theta$, between $0$ and $2\pi$. {Although such uniform distributions are unrealistic, this assumption turns out to be helpful, especially for characterising the purity and completeness of our detection algorithm.} Analogously, the right ascension (RA) and Dec co-ordinates of the SSOs were generated to have at least $N/36$ objects per VIS CCD and to completely encompass the four dither images scheduled for any observing field.

  The VIS magnitudes were sampled in six bins from $20$ to $26$, each one magnitude wide. We note that the saturation limit for a SSO depends on both the magnitude and apparent motion of the object (see discussion below) as well as on, obviously, the instrument electronics. In the following, with the aim of having good enough statistics, we performed the simulation assuming $N_{\rm SSO}=2000$ SSOs per full focal plane, resulting in approximately 50 moving objects per VIS CCD in any bin of SSO magnitude and apparent motion considered.

  For any SSO that fell in a CCD, we computed the number of integrated electrons, $C_{\rm SSO}$ (i.e. accumulated within the exposure time, $t_{\rm exp}$) depending on the object magnitude, $m$, and based on the VIS zero-point ($I_{\rm E}=25.58$) as $C_{\rm SSO}=t_{\rm exp}10^{-(m-I_{\rm E})/2.5}$. Assuming a velocity, $V_{\rm SSO}$ (in units of $\arcsec\,{\rm h}^{-1}$), the streak will be $L_{\rm SSO} \approx V_{\rm SSO} t_{\rm exp}/(3600 p_{\rm fov})$ pixels long so that the object counts per pixel are roughly $C_{\rm SSO}/L_{\rm SSO}$. Hence, we simulated a SSO as a sequence of a number of stars falling in nearby pixels.

  Since, in this scheme, a moving object is simulated as a sequence of stars, an oversampling factor of ten was used to avoid PSF undersampling effects. Finally, we convolved each trail with the instrumental PSF, giving rise to realistic SSO signatures in the simulated image.
  We note that pointing inaccuracy and focal plane distortions have not been simulated in our tests. These effects should not influence the SSO detection, provided that slight changes in the associated co-ordinates are corrected in post-processing analysis (see Sect.~\ref{sec:pipe}).
  
\begin{figure*}%[h]
    \centering
    \includegraphics[width=0.4\textwidth]{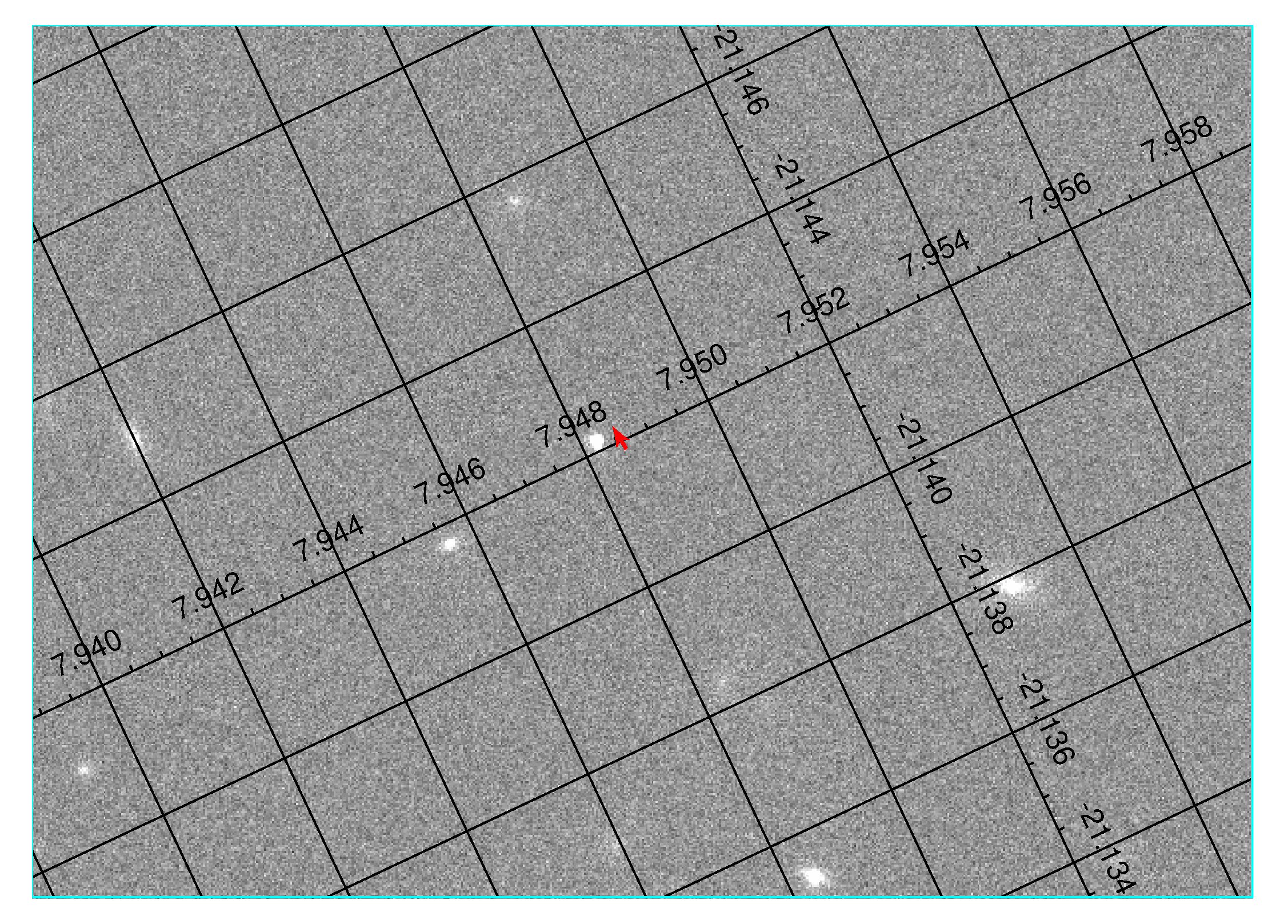}
    \includegraphics[width=0.4\textwidth]{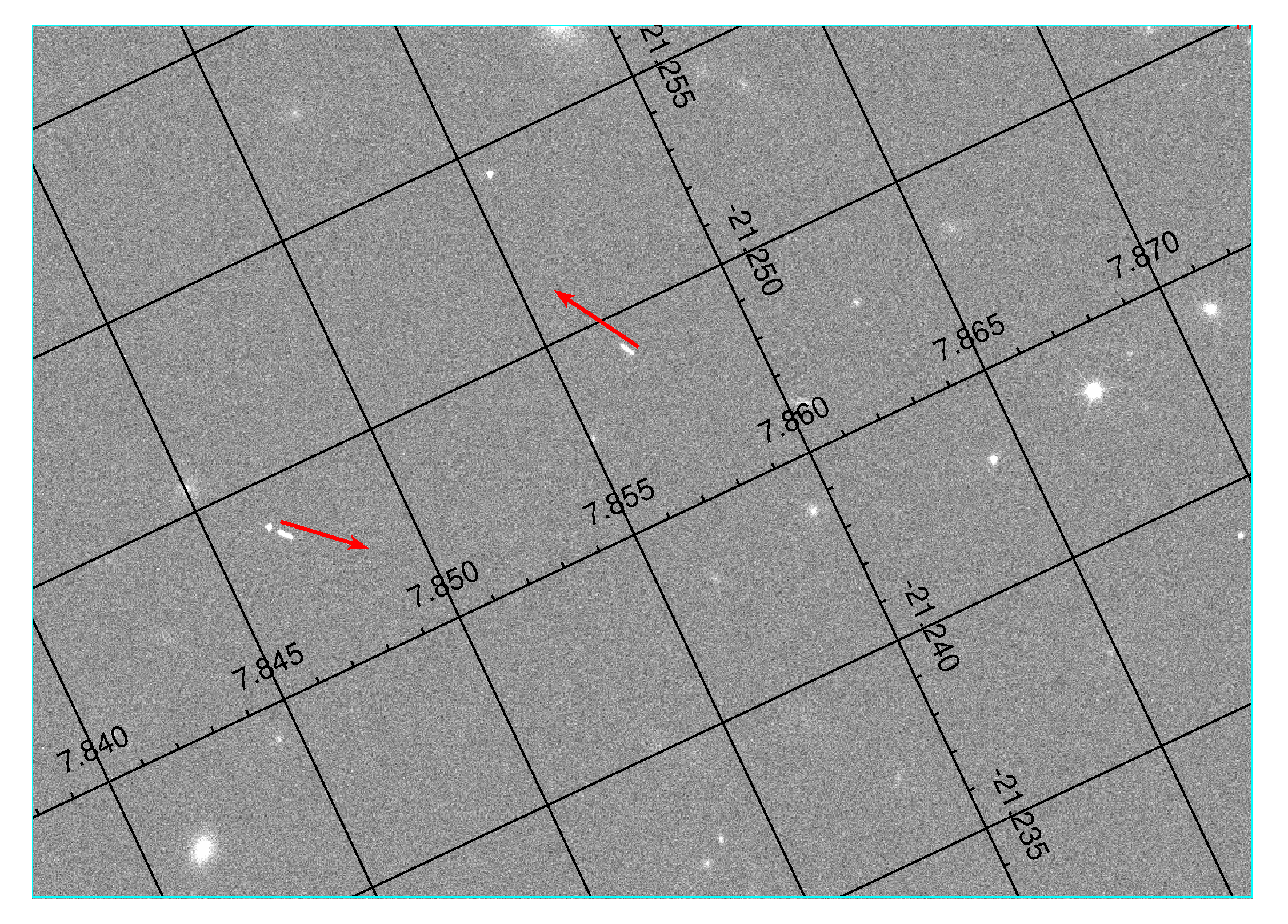}
    \includegraphics[width=0.4\textwidth]{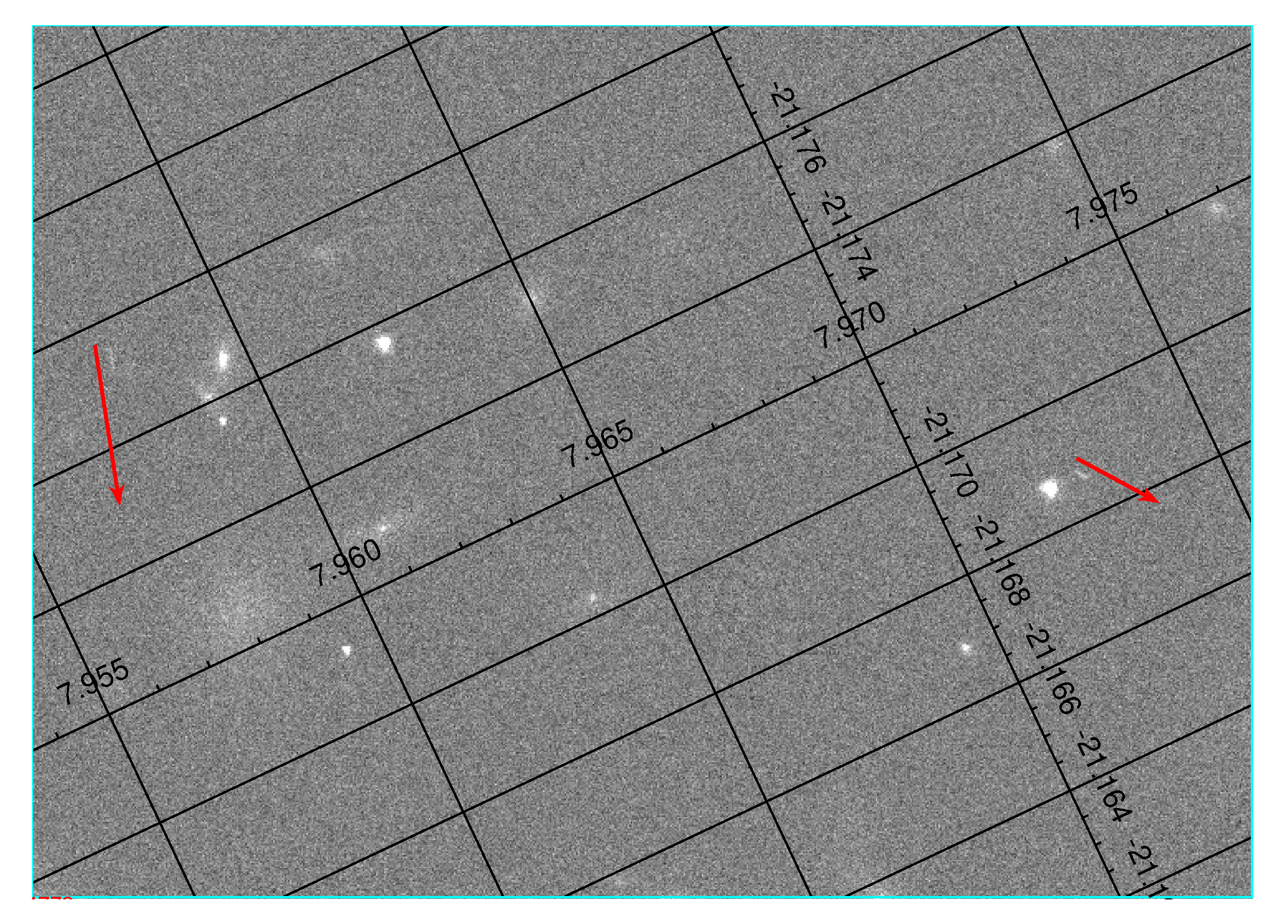}
    \includegraphics[width=0.4\textwidth]{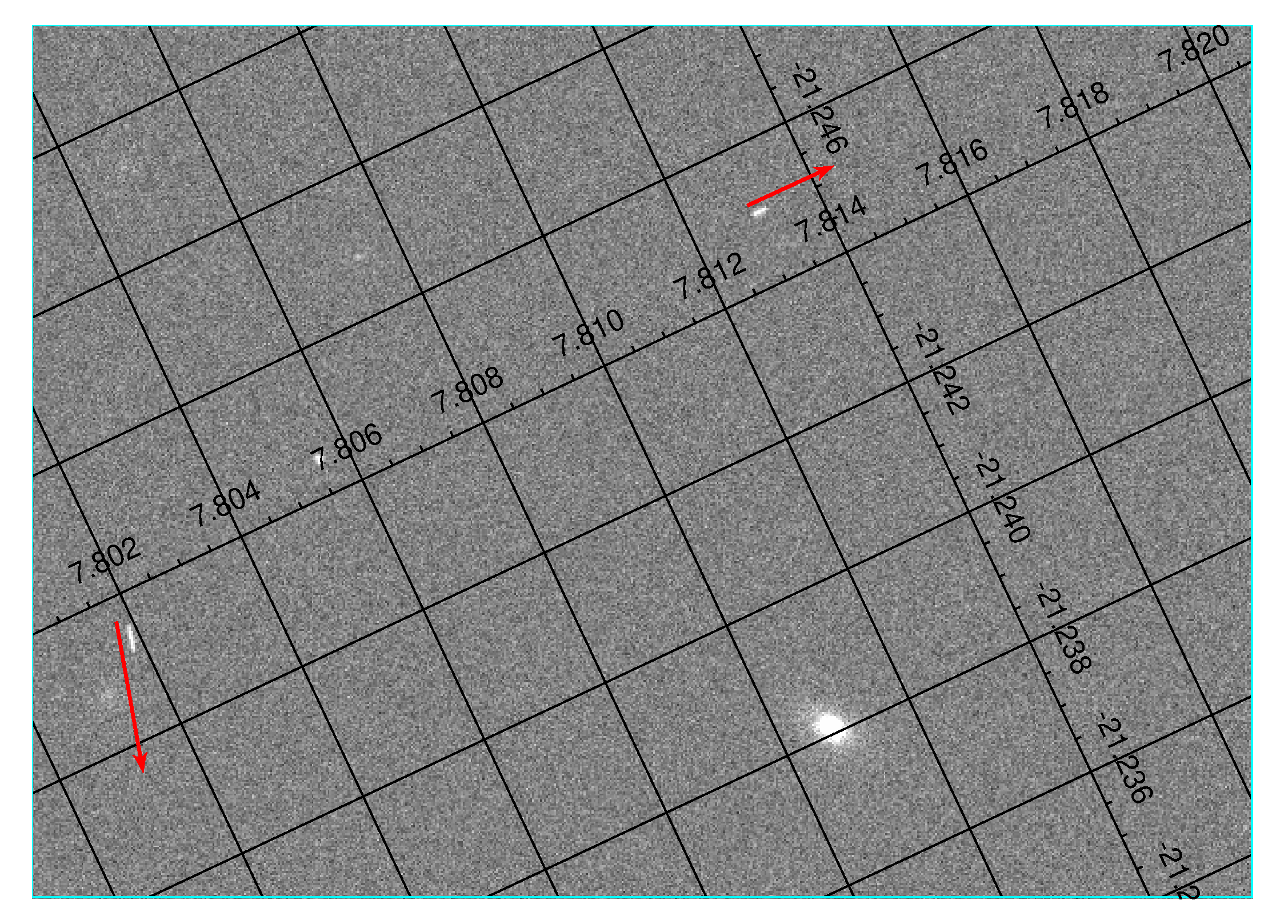}
    \caption{Portions (approximately $\ang{;1.2;}\times\ang{;0.9;}$) of simulated VIS CCDs. The co-ordinate grid shows RA and Dec, while the red arrows indicate the SSO direction in the sky and have lengths proportional to the SSO velocity. From the upper left panel (in a clockwise direction), we show a few simulated objects in the magnitude bins 20--21, 22--23, 23--24, and 24--25, respectively. The SSO velocity is in the range of $0\arcsecf1$--$10\arcsec\,{\rm h}^{-1}$. Note that a small shift has been applied to each arrow to make the SSO streak behind it more evident.}
    \label{fig:simulated_all}
\end{figure*}

{As an example, in Fig.~\ref{fig:simulated_all}, we show small portions of VIS CCDs for one of the simulated dithers. The arrows identify the position of SSOs with a velocity of $0\arcsecf1$--$10\arcsec\,{\rm h}^{-1}$ (the arrow lengths being proportional to the input SSO speed) with magnitudes spanning the range from 20 to 25.
 }

\section{Detection method}
\label{sec:pipe}

  \texttt{SSO-PIPE} was developed to search for objects that appear to move between different exposures and can work directly on images as well as catalogues of objects with the associated (astrometrically corrected) co-ordinates\footnote{ 
  A similar method based on \sext and \texttt{SCAMP} is described in \citet{bouy2013} and \citet{mahlke2018} when searching for sources at different positions in dithered images as those provided by the Kilo-Degree Survey (\url{http://kids.strw.leidenuniv.nl}).}. On the timescale of the typical VIS exposure and between subsequent images, SSOs appear to move with respect to background sources with a velocity that can be as slow as a fraction of an arcsecond per hour. 
  
\begin{figure*}%[h]
    \centering
    \includegraphics[width=\hsize]{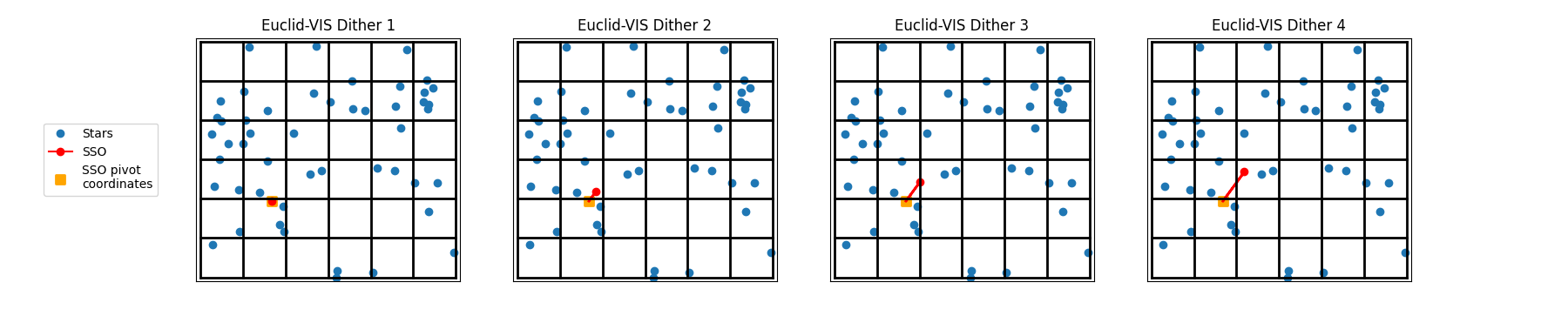}
    \caption{Simplified representation of the {\tt SSOFinder} algorithm at work. 
    A {PIVOT} source (orange dot in dither 1 panel) is considered a good SSO candidate if it appears to move in the remaining images (red dots in subsequent dithers) and all the tracklet members are on the same trajectory (see text).}
    \label{fig:fp_ssofinder}
\end{figure*}

  The method adopted for the blind search of SSOs consists of several steps that can be summarised as follows.
  \begin{enumerate}[topsep=0pt]
    \item{When \texttt{SSO-PIPE} was fed with four VIS observations, it reconstructed the full VIS focal plane and applied master bias, flat, and dark corrections. It then searched for cosmic-ray signatures using the \texttt{L.A. Cosmic} algorithm \citep{lacosmic}, as was implemented in \texttt{Astro-SCRAPPY}\footnote{\url{https://github.com/astropy/astroscrappy}}, consisting of a Laplacian edge-detection filter that identified and removed all bright pixels generated by cosmic rays of arbitrary shapes and sizes.}

    \item{The images were then astrometrically calibrated, which was the most time-consuming part of the pipeline. Starting from the raw images, we performed the astrometric correction using a two-step procedure. First, stellar patterns (asterisms) observed in the analysed field were searched over the available catalogues, namely Gaia DR3 \citep[GDR3,][]{2023A&A...674A...1G}, Pan-STARRS \citep{2018AAS...23110201C}, and unWISE \citep{2019ApJS..240...30S}, and cross-correlation was performed via \texttt{astrometry.net} \citep{2010AJ....139.1782L}. As a second step, a final astrometric adjustment was made using \texttt{SCAMP} \citep{2006ASPC..351..112B}, which also accounts for deformations in the field of view. Here, as was described in \citet{2006ASPC..351..112B}, \texttt{SCAMP} minimises a distance function that depends on the co-ordinates of detected sources matched with objects in the reference catalogue. During this procedure, to prevent any divergence in the astrometric solution, we also required that the relative CCD positions be dictated by the geometry of the VIS detector.}

    \item{We then ran \sext software \citep{bertinsex1} on each image and got four final catalogues of detected sources with an accurate astrometric position. The detection threshold ({\tt DETECT\_THRESH} in \sext) was set to three standard deviations from the local background, and a minimum of four adjacent pixels ({\tt DETECT\_MINAREA}  in \sext) above the noise level was required. We also requested that the minimum contrast in deblending the sources be $0.05$ ({\tt DEBLEND\_MINCONT} in \sext). This requirement comes from the fact that SSOs can often move so fast that they appear as long trails instead of small elongated point-like objects. Due to the inevitable noise in a SSO feature, too small a deblend parameter would lead \sext to split a single SSO trajectory into multiple targets. On the other hand, a much larger value would result in merging close-by-chance sources in a single detection. Sources identified close to CCD gaps and borders or showing broken isophotes were not considered. The values suggested here and set in the final run of the pipeline were derived empirically.}

    \item{We applied our \ssofinder algorithm (detailed below) to \sext catalogues to identify moving-object candidates. This was done by directly comparing the source co-ordinates with those in a reference catalogue (hereinafter, the {PIVOT} catalogue). We required that the target appear in at least {\tt NOBS\_SSO=3} of all the available catalogues.

    A source in the PIVOT was flagged as a potential SSO candidate if no other object was found in the remaining three catalogues within a minimum distance {\tt MIN\_DIST} (expressed in arcseconds). Then we searched for tracklets; that is, ensembles of sources within {\tt MIN\_DIST} and {\tt MAX\_DIST}, the latter value being the maximum distance travelled by a SSO in a given time,
    \begin{equation}
         \Delta t_{\rm trv} = 4t_{\rm exp}+3t_{\rm step}\;,
    \end{equation}
    where $t_{\rm step}$ is the typical time between the end of a data acquisition and the start of the following one. The maximum speed considered was $10\arcsec\,{\rm h}^{-1}$.  
    Figure~\ref{fig:fp_ssofinder} sketches how the code works. In particular, for a given {PIVOT} SSO candidate (orange dot), the algorithm finds associations with nearby sources appearing in the remaining catalogues and estimates (for each target pair) the velocity, 
    \begin{equation}
        \mu = \sqrt{\mu_\alpha^2 + \mu_\delta^2}\;.
    \end{equation}
    Here, the proper motion components ($\mu_\alpha$ and $\mu_\delta$) along the RA and Dec axes are given by
    \begin{equation}
        \mu_\alpha = \frac{\Delta\alpha}{\Delta t} \> \cos{\delta}\;,
        \\
        \mu_\delta = \frac{\Delta\delta}{\Delta t}\;,
    \end{equation}
    where $\Delta\alpha$, $\Delta\delta$, and $\Delta t$ are the differences in RA, Dec, and time between each of the considered pairs of sources, respectively. For each pair of sources, the direction of motion can be estimated by evaluating the angle,
    \begin{equation}
        \theta = \arctan{\frac{\Delta\delta}{\Delta\alpha}}\;.
    \end{equation}
    Different entries in the catalogues are associated with the same tracklet if their velocity (with respect to the PIVOT source) is constant and the tracklet members lie in a straight line. A candidate qualifies as a detected object when the evaluated velocity and position angle remain constant within the fixed errors for the proper motion ({\tt ERR\_PROPMOT}) and position angle ({\tt ERR\_POSANGLE}).
    
    Finally, for any selected candidate that fulfils the above conditions, the average values for the proper motion and position angle are evaluated as 
    \begin{equation}
        \overline{\mu} = \frac{1}{N} \sum_{i} \mu_{\rm P,\it i}\;,
        \\
        \overline{\theta} = \frac{1}{N} \sum_{i} \theta_{\rm P, \it i}\;,
    \end{equation}
    where {\rm P} indicates the {PIVOT} catalogue and $i$ the $i$-th catalogue of the series.

    With the above scheme, the algorithm fails to recover SSO candidates that are not imaged onto the focal plane at the time of the first dither image (and that are not in the associated catalogue either) but that appear only in the subsequent exposures. Therefore, a second run is required, using the second dither image as the {PIVOT} reference. Of course, duplicate candidates must be removed from the final merged catalogue. We require at least three detections for a moving object to be flagged as a potential SSO candidate. Therefore, since the \euc dither pattern comprises four images, running the search algorithm twice is enough to cover all the possibilities.}
  \end{enumerate}

  All the sources that satisfy the previous criteria are flagged as possible SSOs, and an alert containing the object co-ordinates and the associated expected proper motion is raised. For each detected SSO target, a stamp image of 101\,$\times$\,101 pixels 
  centred on the source is also given for inspection purposes (see also Sect.~\ref{section:som}).
  
  In the following, we present an analysis of the pipeline efficiency by studying the purity and completeness of the recovered SSO sample against the simulated one in different bins of apparent magnitude and apparent motion.
  %Finally, we remind the possibility of skipping steps (1.) through (4.) in case the final four catalogues rather than images are in input to SSO-PIPE. In this case, only step (5.) is necessary.

\section{Purity and completeness of the detection tool}
\label{sec:disc}

  We assessed the behaviour of the \texttt{SSO-PIPE} pipeline by determining the purity and completeness of the output samples. In this respect, a sample was considered pure if objects detected as SSOs consisted of genuine simulated moving sources against the ‘detection background’ of stars, galaxies, and spurious features erroneously identified as SSOs by the algorithms. Therefore, we define (in each bin of width ${\rm d}v$ centred at $v$) the purity of the sample as 
  \begin{equation}
    {\rm Purity}(v) = {\rm NDA}(v)/{\rm ND}(v)\;,
    \label{purity}
  \end{equation}
  where ${\rm ND(v)}$ is the number of objects detected as moving sources in each bin of velocity, and ${\rm NDA}(v)$ is the number of genuine associations; that is, those objects identified as moving sources corresponding to real SSOs in the simulated input catalogue.

  Similarly, the completeness of the output sample can be defined as the number of genuine associations, ${\rm NDA}(v)$, over the number of all simulated SSOs, ${\rm N}(v)$, in a particular velocity bin as
  \begin{equation}
    {\rm Completeness} (v) = {\rm NDA}(v)/{\rm N}(v)\;,
    \label{completeness}
  \end{equation}
  so that, ideally, one would expect both purity and completeness close to unity for a perfect algorithm.

 Of course, identifying a stationary source (star, galaxy, and spurious CCD artefact) as a SSO or assigning the detection to the wrong velocity bin (thus decreasing both purity and completeness) depends mainly on the random pixel noise, the CCD astrometric solution (that affects the low-velocity bins for faint objects), and the detection parameters, ${\tt MIN\_DIST}$ and ${\tt MAX\_DIST}$, in \ssofinder, fixed here as \ang{;;0.04} and 12\arcsec, respectively. In particular, the ${\tt MIN\_DIST}$ value was chosen as the typical astrometric accuracy in \euc VIS images. On the other hand, a larger value of the ${\tt MAX\_DIST}$ parameter resulted in many fake detections in the case of large velocity bins due to the faintness of the source.

 After testing, we found a combination of parameters that appears to maximise the purity and completeness in the apparent motion range, $0\arcsecf1$--$10\arcsec\,{\rm h}^{-1}$, of the recovered SSO sample. There is a trade-off between maximising the number of SSO detections (true-positive targets) and minimising the number of fake moving sources recognised erroneously (false positives) by the pipeline.     
\begin{figure*}
    \centering
    \includegraphics[width=0.5\textwidth]{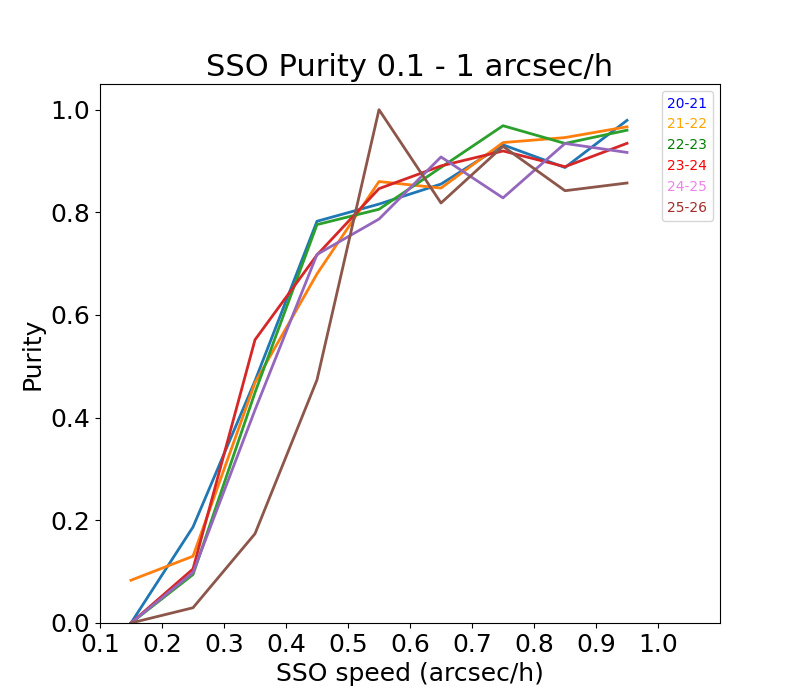}%
    \includegraphics[width=0.5\textwidth]{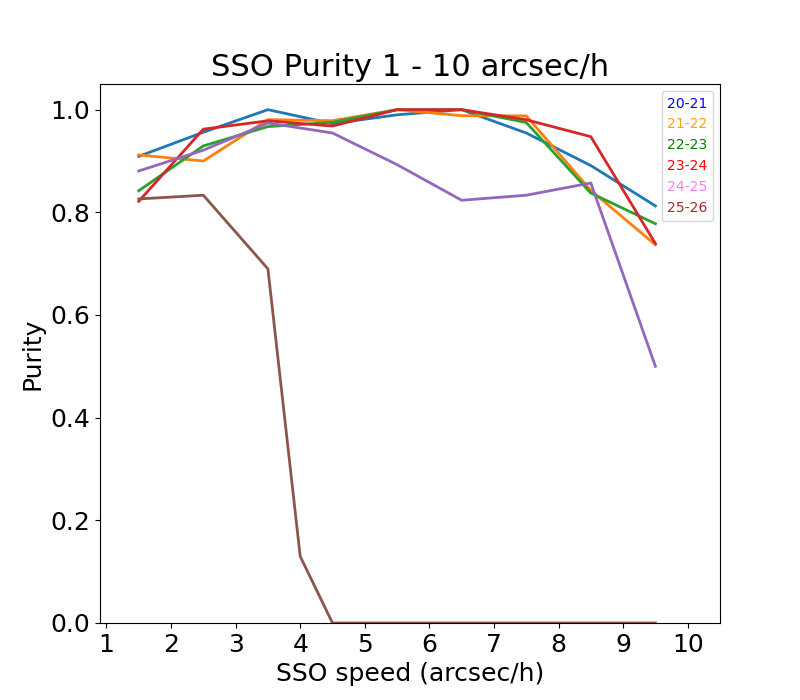}
    \caption{Purity of the SSO sample recovered by the pipeline (see text for details) as a function of the object velocity and for different magnitude bins. Note that the purity abruptly decreases for the faintest SSOs (in the bin magnitude of 25--26) with an apparent motion faster than $4\arcsec\,{\rm h}^{-1}$.}%
    \label{fig:purity}%
\end{figure*}

\begin{figure*}
    \centering
    \includegraphics[width=0.5\textwidth]{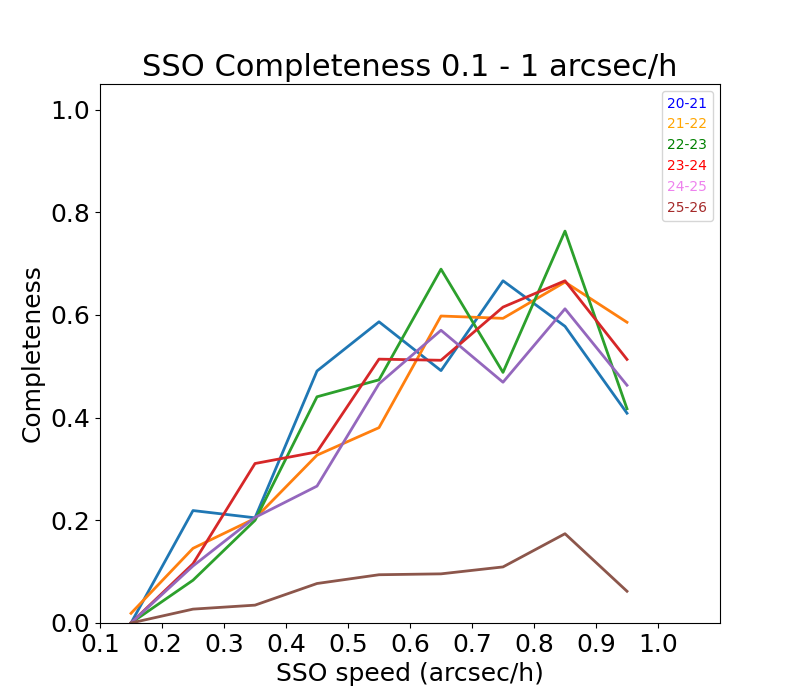}%
    \includegraphics[width=0.5\textwidth]{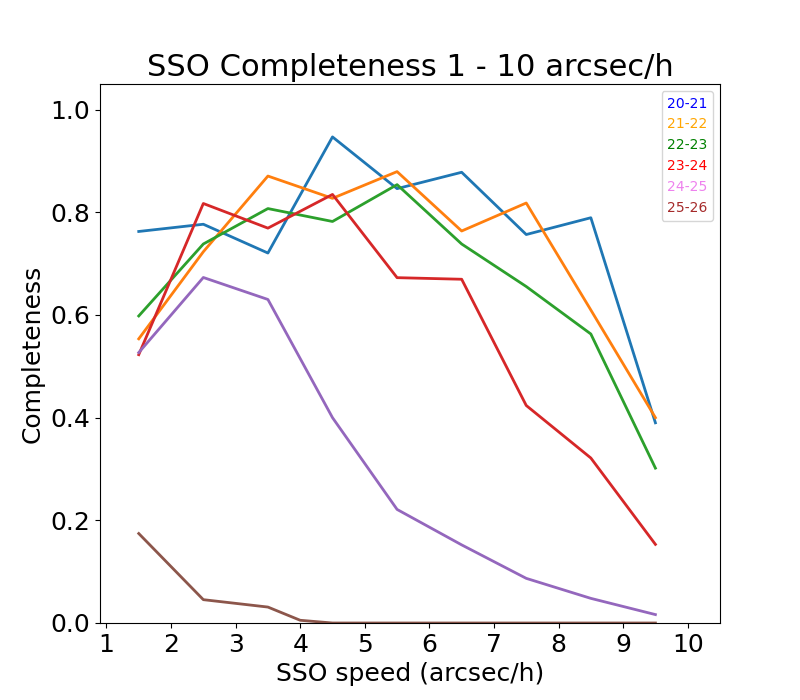}
    \caption{Completeness of the SSO sample as a function of velocity and for different magnitude bins.}%
    \label{fig:completeness}%
\end{figure*}

 In Fig.~\ref{fig:purity} and \ref{fig:completeness}, we show the purity and completeness of the SSO sample recovered by the pipeline as a function of the asteroid's apparent motion and for different magnitude bins. In the range of $1\arcsec$--$10\arcsec\,{\rm h}^{-1}$ (right panel), purity remains close to a $90\%$ level for all the selected magnitude bins apart from the faintest one (25--26\,mag), where, as is expected, it abruptly decreases to zero for SSOs with a velocity larger than $4\arcsec\,{\rm h}^{-1}$. On the other hand, completeness remains of the order of $80\%$ for SSO speeds in the range of $1\arcsec$--$10\arcsec\,{\rm h}^{-1}$ and for the brighter bins of magnitude, but it rapidly falls for higher velocities and fainter objects.
 
 The main reason for the lower completeness of faster SSOs in the bins of higher magnitudes is that fast SSOs are fainter. Assuming a SSO with a magnitude of 25.5, the expected total count number in the $I_{\rm E}$ band is approximately $610$. For an assumed velocity of $5\arcsec\,{\rm h}^{-1}$, the SSO produces only approximately $80$ counts per pixel, which, for a zodiacal background as in \citep{scaramella2022}, corresponds to $4\,\sigma$ over the background. Furthermore, a decrease in purity is also visible for velocities larger than $9\arcsec\,{\rm h}^{-1}$ and, on the other side of the inspected range in apparent motion, below $0\arcsecf4\,{\rm h}^{-1}$, a large decrease for all bins of magnitude comes up as well. In both cases, this is due to the intrinsic scatter that affects the centroid detection algorithm, which decreases the purity as more fake targets enter the sample.

\section{Classification of Solar System objects based on self-organising maps}
\label{section:som}

  Once any detection tool is run on an image and a catalogue of SSO candidates is produced, one still needs to assess the quality of each candidate. Due to the large \Euclid data volume, a visual inspection of each SSO candidate is impractical. Therefore, one must rely on a machine-learning approach to analyse the problem. In this respect, machine-learning techniques and neural networks have already proved to be very useful in astronomy with respect to, for example, the classification of different galaxy morphology types \citep{Odewahn, Dieleman}, stellar spectra classification \citep{gulati}, and the detection of strong gravitational lensing arcs \citep{Schaefer}. \citet{lieu2019} trained deep CNNs on simulated VIS images to classify SSOs based on morphological properties. The CNNs were not used to detect the SSOs (a list of potential SSOs was, in fact, provided to the network, \citealt{lieu2019}) but to reject false-positive detections. The CNNs separated the SSOs from other astronomical sources with an efficiency of $96\%$ for apparent velocities larger than $10\arcsec\,{\rm h}^{-1}$ down to magnitude 26. \citet{2023A&A...679A.135P} used a CNN to detect streaks and their co-ordinates in \euc images and then a recurrent neural network to merge long streaks recognised as multiple targets by the previous step. The authors thus show the possibility of improving the detection efficiency of asteroid streaks using deep learning, especially for faint objects not detected by other methods.

  With the ultimate goal of giving a classification for the objects detected by any deterministic algorithm (such as the one presented in the previous Sections, and for SSOs with velocities lower than $10\arcsec\,{\rm h}^{-1}$ down to magnitude 26), we present an algorithm designed to classify sources (stars, galaxies, and SSOs) based on their un-parameterised images. The method uses a form of neural network; namely, a SOM  \citep{Kohonen}. Due to the simplicity of the algorithm implementation and the capability to identify unknown features in the data, SOMs are widely used in astronomy; for example, in star and galaxy classification \citep{mahonen, miller}, galaxy morphology classification \citep{naim, molinari}, classification of gamma-ray bursts \citep{rajaniemi}, studies of mono-periodic light curves \citep{brett}, and studies related to the calibration of photometric redshifts observed by \euc \citep{masters, saglia2022}. We show here how to build an SOM architecture, train it, and extend its functionality to the classification of new data sets.

\subsection{Preparing the data for self-organising-map training}
\label{sec:somtrain}

  In order to test the capability of an SOM to classify objects in an unsupervised fashion, we dealt with the simulation of thousands of different objects. In particular, to classify stars, galaxies, and SSOs, we concentrated on simulating stamps, each containing a specific source. We modelled each object of interest with a particular set of parameters such as {the magnitude}, luminosity profile, ellipticity, and position angle (for a galaxy), as well as velocity and direction of motion (for a moving SSO). We then built catalogues for training and testing the model (see Sect.~\ref{sec:som-1}--\ref{sec:som-3}). Each catalogue consists of 4000 random images in the form of 101\,$\times$\,101-pixel matrices (suitable to host slow-moving SSOs), and the pixel size is that of a VIS image; that is, $0\arcsecf1\,{\rm pixel}^{-1}$. 
  
  \begin{figure*}
    \centering
    \includegraphics[width=0.8\textwidth]{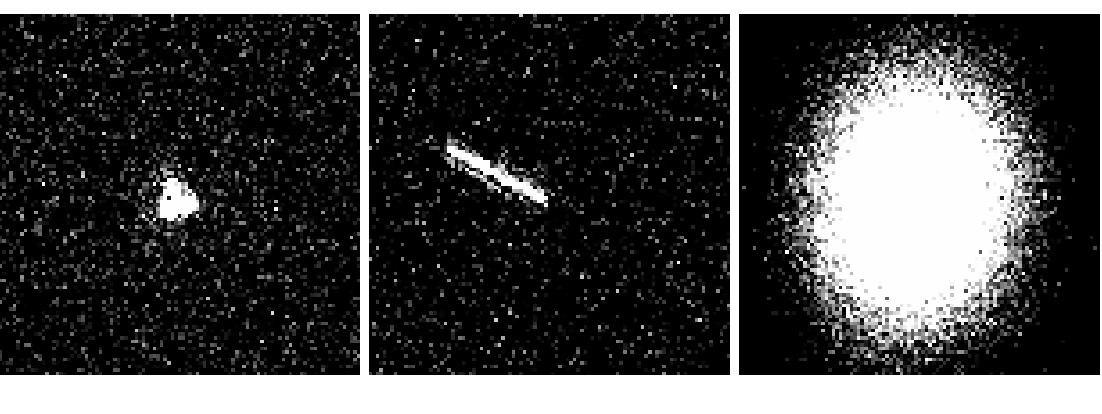}%
    \caption{Example of simulated star (left), SSO (middle), and galaxy (right). The data are normalised, as is described in the text.
    } 
    \label{exampledata}%
  \end{figure*}
  
  Any particular image can host an object of the above type with stars and galaxies simulated with the \texttt{Galsim} software \citep{galsim}. In particular, stellar objects were simulated as point-like sources, while galaxies were assumed to follow a Sérsic surface brightness profile characterised by three properties: the Sérsic index, $n$, the integrated flux (derived from the object magnitude, the VIS zero-point, and exposure time), and the half-light radius, $r_{\rm e}$. For each galaxy, given these properties, the surface brightness profile scales with respect to $r$ as 
  \begin{equation}
    I(r)\propto \exp[{-b(r/r_{\rm e})^{1/n}}]\;,
  \end{equation}
  where $b$ is constrained to give the correct $r_{\rm e}$ value. Furthermore, each galaxy was deformed to account for an associated ellipticity and orientated in the sky according to a random position angle between $0$ and $360$ degrees. For each galaxy, we uniformly extracted a half-light radius in the range of $1\arcsec$--$3$\arcsec (so that each stamp size is a factor of    $2.5$ larger than the maximum half-light diameter) and an ellipticity value between $0$ and $0.8$. {The Sérsic index, $n$, was selected randomly in the range  between $0.5$ and $4$, the latter corresponding to the de Vaucouleurs galaxy profile. The integrated galaxy magnitude was uniformly selected in the range of $14$--$24$.
  {It should be noted that training the neural network with galaxies that have Sérsic indices in the above range would allow one to correctly classify such objects (see Sect. \ref{somprob-section} for a further discussion), but would fail for galaxies with different luminosity profiles and/or with more more complicated shapes, since the SOM noise enhances. 
  }}{Although the uniform distributions for the relevant parameters are unrealistic, this assumption is required to avoid biases during the SOM training phase and, furthermore, allows one to evaluate the completeness and purity of the classification regardless 
  of the distribution details.}

  Once a star or a galaxy object has been simulated, the resulting image was then convolved with the VIS PSF and Poisson noise was added at a level of the noise expected in the data. Each stamp was also characterised by the same bias level (in counts per pixel) and a random read-out noise at a level of 3.5 counts per pixel. Knowing the exact value of the bias level is not crucial here, since each data set is first normalised before exposing the neural network to it.

  We simulated the SSOs as a sequence of point-like objects (oversampling by a factor of ten to avoid PSF undersampling effects) and then convolved them with the instrumental PSF, as is described in Sect.~\ref{sec:simu}. This procedure results in realistic SSO signatures in the simulated image. Also, in the case of the SSO simulation, we used the intrinsic features of the Galsim software to position the target onto the final image and for the PSF convolution. In Fig.~\ref{exampledata}, just as an example, we show a test data sample for a simulated star, SSO, and galaxy, respectively.

    {
Finally, we stress again that, although Sérsic models reproduce the overall surface brightness of galaxies well (see e.g. \citealt{peng2002}), the spatial resolution of the VIS images allows one to capture detailed features (such as bulges, spiral arms, and knots) in many of the observed galaxies, increasing the shape complexity. However, the increasing shape complexity acts against the object classification so that the intrinsic noise of the classifier would increase.
  We also remind the reader that the training sample is such that the relative weight among the simulated objects is equal. In other words, approximately a third of the training samples fall into each class of stars, SSOs, or galaxies. Since a Kohonen SOM classifies one image at a time, we require that each stamp contain only one specific object \citep[for a similar case study, see e.g. ][]{bretonniere}. Object blending (images that might contain multiple categories amongst those searched for) increases the SOM noise against the target retrieval. 

  Quantifying the effects of resolved structures on the SOM image classifier as well as the impact of source blending on the  SOM training and forecast is outside the purpose of the paper and will be investigated in future work. In any case, any SOM would need to be retrained with real \euc data before offering reliable classifications of real detections.
  }

\subsection{Building a self-organising map}
\label{sec:som}

  An SOM consists of a set of $Q=N\,M$ neurons or nodes, typically organised in the form of a rectangular lattice. Each neuron of co-ordinates $(i,j)$ (with $i=0,..., N-1$ and $j=0,...,M-1$) is characterised by a set of $K$ values representing the components ($k=0,...,K-1$) of the ‘reference vector’ associated with that particular neuron. Each reference vector (with components $r_k$) of the map is exposed to an input vector (each of which has components, $w_k$, i.e. the same $K$ cardinality as the reference vectors) taken from a training data set consisting of $L$ inputs, such as images. The main purpose of the SOM training is to detect common patterns amongst the $L$ inputs so that the data are ideally divided into well-separated groups. The number of $Q$ neurons in the SOM follows from requiring that the resulting map not be too large to have one single neuron adapted per input data (the main goal is to generate clusters of similar data). Similarly, a map that is too poor fails to catch an adequate organisation of the data into separate classes. A common practice for a squared SOM ($N=M$) is to select a map size of $N=(5\sqrt{L})^{1/2}$. We verified that a squared SOM with $N=20$ neurons per side is sufficient to classify the data in our sample and to account for the appearance of stellar, galaxy, and SSO clusters in the map.

  The SOM is trained by exposing all the neurons of the map iteratively to each sample in the training data set and by determining, for each sample, the associated winning neuron. We say that an ‘epoch’, $t$, passes when the SOM processes all the $L$ inputs once. The final goal of training the network is to modify the values of the reference vector components so that the training samples that show some similarity are placed in nearby neurons. After each sample is associated with a node, the weights of the best winning neuron (and its close neighbours) are updated. The weight update continues until all the samples are passed to the SOM and the next epoch starts. Then, the entire procedure is repeated until the training members settle into (self-organised) clusters.

\subsection{Determining the winning neuron}
\label{sec:som-1}
  
  As is described above, the SOM is trained by passing, at a given epoch, $t$, the whole training set to the network, and determining, for each data sample, the best matching neuron minimising some distance function. A validation set is also passed to the neural network (see Sect.~\ref{sec:som-3}) to evaluate the SOM performance.

  Here, the co-ordinates $(i^l_{\rm win}, j^l_{\rm win})$ of the winning node for an $l$-th input data sample (with $l=0, ..., L-1)$ are those obtained by minimising the Euclidean distance between the input vector and each reference vector [associated with the $(i, j)$ pixel] in turn; that is, the quantity
  \begin{equation}
    D^l_{\rm min} = {\rm min}_{i,j} \left( \sqrt{\sum _{k=0} ^{K-1} m_k^l(r_k^{i,j}-w_k^{l})^2} \right)\;,
  \label{euclid_distance}
  \end{equation}
  where the factor, $m_k^l$, is a mask that accounts for any missing value (or NaNs) in the input vector, $l$. In particular, by requiring that the mask equals one for any existing input vector component and zero otherwise (see e.g. \citealt{missom}), the SOM algorithm can easily handle missing elements in the data sample.

  Once a neuron of co-ordinates $(i, j)$ has been flagged as the best matching node for an input sample, $l$, the $k$ components (weights) of the node are updated together with the weights of close neurons according to the rule
  \begin{equation}
    r'^{i,j}_k = r^{i,j}_k+\alpha(t/N_{\rm e}) H(t/N_{\rm e},{\bf d}_{\rm win}-{\bf d})\left(w_k^l-r_k^{i,j}\right)\;,
  \label{ww1}
  \end{equation}
  where $\alpha(t/N_{\rm e})$ is the learning rate coefficient, $H(t/N_{\rm e}, {\bf d}) $ is the neighbourhood updating function, and ${\bf d}_{\rm win}$ and ${\bf d}$ indicate the vector positions of the best winning neuron and a nearby node, respectively. Here, we adopted a Gaussian function of the form 
  \begin{equation}
    H(t/N_{\rm e}, {\bf d}) =\exp\{{-d^2/[2\sigma^2(t/N_{\rm e})]}\}\;.
  \label{ww2}
  \end{equation}
  The time co-ordinate, $t$, varies linearly with the epoch number, from $t = 0$ (at the first iteration) to $t=N_{\rm e}$ ($N_{\rm e}$ is the number of epochs). Moreover, the presence of a smooth neighbourhood kernel function, characterised by a variance $\sigma^2(t)$, enables the formation of clusters of nodes capable of catching similarities in the data.

  The convergence of an SOM towards a stable configuration depends (at each epoch, $t$) on the learning rate, $\alpha(t)$, which drives the blending of the reference vectors, and the $\sigma(t)$ parameter, which affects the number of neurons (close to the winning node) whose weights are updated (according to Eqs.~\ref{ww1}--\ref{ww2}) after each input sample is passed to the SOM. We studied the effects of varying $\alpha(t)$ and $\sigma(t)$ according to two possible different decreasing functions of time, $t$: a linearly decreasing monotonic function and an exponentially decreasing one. In the linear case, the functional form is
  \begin{equation}
    p(t)=p_0\,(1-t/N_{\rm e})\;.
  \label{linear}
  \end{equation}
  Analogously, the exponentially decreasing function reads out to be
  \begin{equation}
    p(t)=p_0\,\exp(-\lambda t/N_{\rm e})\;,
  \label{exponential}
  \end{equation}
  where, in both cases, $p(t)$ is either $\alpha(t)$ or $\sigma(t)$, and $p_0$ gives the associated starting value of the involved quantity. In the exponential case, $\lambda$ represents a scale parameter so that $p(t)=10^{-3}p_0$ at the last iteration. The above scheme ensures that large-scale structures form in the map at a very early stage of the training procedure, and then they become stable (with little changes) at late epochs.

\subsection{Monitoring the self-organising map and early stopping of the training}
\label{sec:som-3}

  The learning behaviour of the SOM can be evaluated by using a statistic indicating the difference between the input samples (input vectors) and the reference vectors associated with each neuron. At each epoch, we evaluated the minimum distance between a sample and the associated winning neuron reference vector (see Eq.~\ref{euclid_distance}) averaged over all members of the training set as
  \begin{equation}
    S(t)=\frac{\sum_{l=0}^{L-1} D^l_{\rm min}}{L}\;.
  \label{euclid_distance_average}
  \end{equation}
  
  We expect that the SOM error, $S(t)$, turns out to be relatively large at the early stages of the training process and decreases as the number of iterations increases. After the SOM changes the learning rate and variance values ($\alpha(t)$ and $\sigma(t)$, respectively), it reaches a stable condition, and the average error shows a slower decline. We also expect that, during the initial epochs, the training samples jump in the SOM map so that the winning neuron changes with time. Once the SOM becomes stable, the number of jumps decreases as clusters of similar data form. Therefore, we can define (in analogy to \citealt{brett}) a second measure of the SOM quality; namely, $N_{\rm move}$, which consists of counting the number of training set members that have changed locations on the map in each epoch. We expect that $N_{\rm move}$ is very large at the beginning of the training, because we assigned all the data members to a virtual neuron outside the SOM map. Therefore, in the first epoch, all $L$ members jump to a new position and $N_{\rm move}=L$. The number of movements then decreases, reaches a maximum (at the stage of maximum learning), and then approaches zero when all the clusters of similar patterns have formed.

  From the SOM training scheme described above, it should be clear that $N_{\rm e}$ can be set to a large number, implying a very long, time-consuming procedure. A second effect is that iterating the training over the same data pushes the SOM to an over-fitting state so that the SOM error $S(t)$ decreases (eventually becoming constant when $N_{\rm move}$ approaches zero) and all the weights are exactly adapted to the particular set under examination, but do not generalise for other data.

  Therefore, with the training scheme of the SOM, we decided to implement an ‘early stopping’ method consisting of training the neural network on the input data set and checking the quality of the SOM by evaluating $S(t)$ and $N_{\rm move}$ at each step, but also testing its performance on an independent, randomly chosen validation data set. We then evaluated the SOM error, $S_{\rm val}(t)$, for the validation data and compared it with the SOM error under training. The idea behind the early stopping method is that over the epochs, $S(t)$ and $S_{\rm val}(t)$ both decrease, and when the over-fitting stage is reached, $S_{\rm val}(t)$ shows a minimum and then flattens or starts to increase. We flag this stage as the $t_{\rm over-fitting}$ epoch, stop the learning process, and freeze the weights stored in each neuron. Once trained in this way, the SOM acquires the capability to classify new data.

\subsection{U-matrix associated with the trained self-organising map}
\label{sec:umatrix}

  When the training process of the SOM has been completed, the map can be inspected and searched for clusters of similar neurons. This can be done, for example, by studying the distribution of the training set members that fall on a given node. Here, we prefer to adopt the approach described in \citet{umatrix}, calculating the U-matrix associated with the SOM. A U-matrix simply evaluates the rate of change in the neuron response (the reference vector) across the SOM so that neurons belonging to the same cluster have similar reference vector components, while at the boundaries among clusters, the differences increase. Therefore, for each neuron of the map, we calculate the quantity
  \begin{equation}
    U_{i,j}=\sum _{n=i-1}^{i+1}\sum _{m=j-1}^{j+1}\sum_{k=0}^{K-1}\left(r_{i,j}^k-r_{h,k}^k\right)^2/8\;,
  \label{umatrix}
  \end{equation}
  where, when possible, the eight nearest neighbours of the neuron under investigation are averaged.

\section{Application of an SOM to the classification of stars, galaxies, and Solar System objects}
\label{sec:somend}

\subsection{Training the self-organising map}

  In Sect.~\ref{sec:somtrain}, we described the method adopted to simulate a sample of image data corresponding to point-like stars, galaxies, and SSOs, a few examples of which are given in Fig.~\ref{exampledata}. Each image, simulated initially as a 101\,$\times$\, 101-pixel matrix, is first rebinned by a factor of $f=3$, so that enough details are available for the SOM to determine common patterns, and then normalised to unit variance. Following the training scheme outlined in Sect.~\ref{sec:som}, we found that a map of 20\,$\times$\,20 neurons was capable of autonomously identifying clusters for the three classes of objects hidden in the training data.

  We fixed the starting value of the dispersion parameter, $\sigma(t)$, to four, allowing the SOM to form large clusters at the beginning of the iterations. We tested the behaviour of the network for two initial values of the learning parameter, $\alpha(t)$, namely 0.1 and 0.5, and for each combination of parameters we trained the SOM by assuming the linear or the exponential functions (see Eqs.~\ref{linear}--\ref{exponential}). We always set $N_{\rm e}=500$, and by passing each data sample to the SOM, the winning neuron was identified and the map weights updated.

  In each epoch (once the SOM was exposed to the entire data set), we monitored the performance of the neural network (as is described in Sect.~\ref{sec:som-3}) and updated the values of $\sigma(t)$ and $\alpha(t)$. Consequently, the early-stopping method prevented the SOM from over-fitting to the input data.

  In the proposed scheme, the 20\,$\times$\,20 neurons' reference vectors should be initialised by picking random data from the training set. To test the neural network performance, we always initialised the map with the same weights and selected a combination of starting values for $\sigma(t)$ and $\alpha(t)$, and selected one decreasing function between Eqs.~(\ref{linear}) and (\ref{exponential}).
  
  \begin{figure*}
    \includegraphics[width=0.9\textwidth]{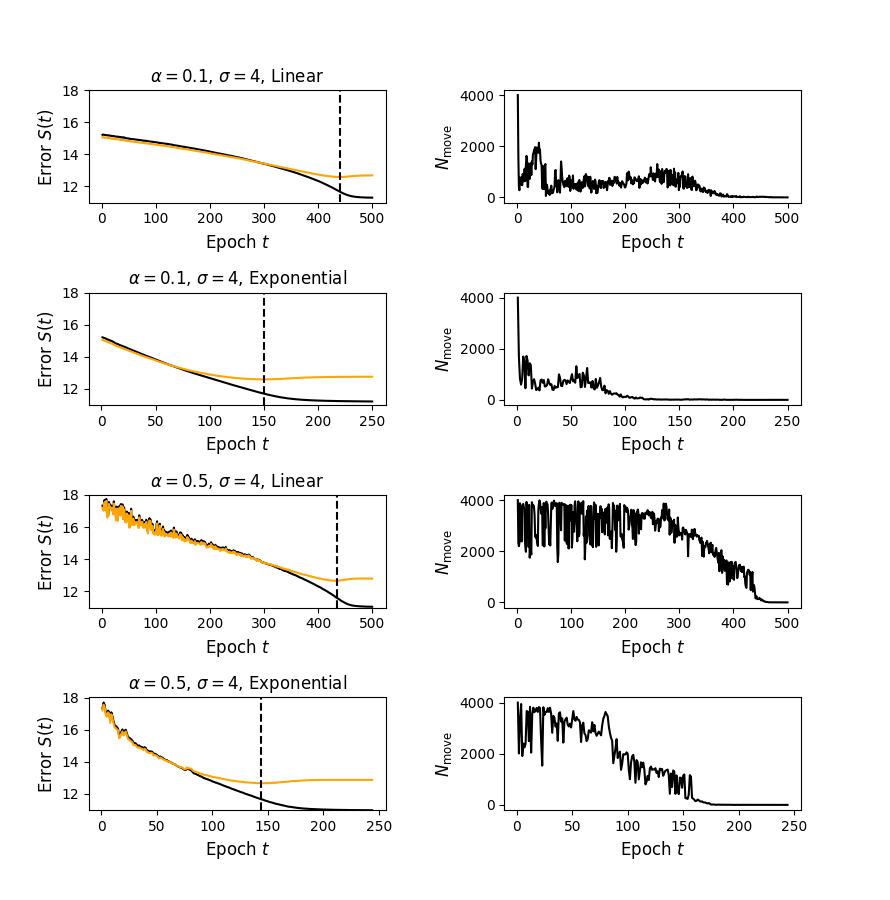}%
    \caption{Performance of the SOM evaluated by calculating the average map error, $S(t)$ (solid black line, \textit{left} panels), and number of movements, $N_{\rm move}(t)$, in a given epoch, $t$ (\textit{right} panels). In the \textit{left} panels, we also give the average error of the SOM when applied to the validation input data (yellow line). The dashed vertical lines represent, for each panel, the epoch associated with the early-stopping method (see text for details).
    %\BC{could you use differnt symbols please? for printed version in BW or color-blind people.
    %}\BC{put a label to the color scale}
    }
    \label{result_1}%
  \end{figure*}

  In Fig.~\ref{result_1}, we show the performance of the neural network by evaluating the error, $S(t)$, and the number of movements, $N_{\rm move}(t)$, at any epoch, $t$. In the left panels, the orange line gives the error, $S_{\rm val}(t)$, evaluated by applying the SOM on a validation data set in order to identify the over-fitting epoch (see Sect.~\ref{sec:som-3} for details), which is flagged by the vertical dashed line. In the first and second rows, the SOM was trained by assuming a starting learning parameter, $\alpha=0.1$, and adopting a linear and exponential decreasing function, respectively. Analogously, in the third and fourth rows, we used $\alpha=0.5$ and again tested the network behaviour with the two adopted decay functions.

  As is evident from Fig.~\ref{result_1}, the result does not depend critically on the starting value of the learning parameter nor the adopted form of the decreasing function. At the end of the last epoch, the training of the SOM with the input data reaches a stable configuration characterised by errors of the same order of magnitude for all runs.

  \begin{figure*}
    \includegraphics[width=1.0\textwidth]{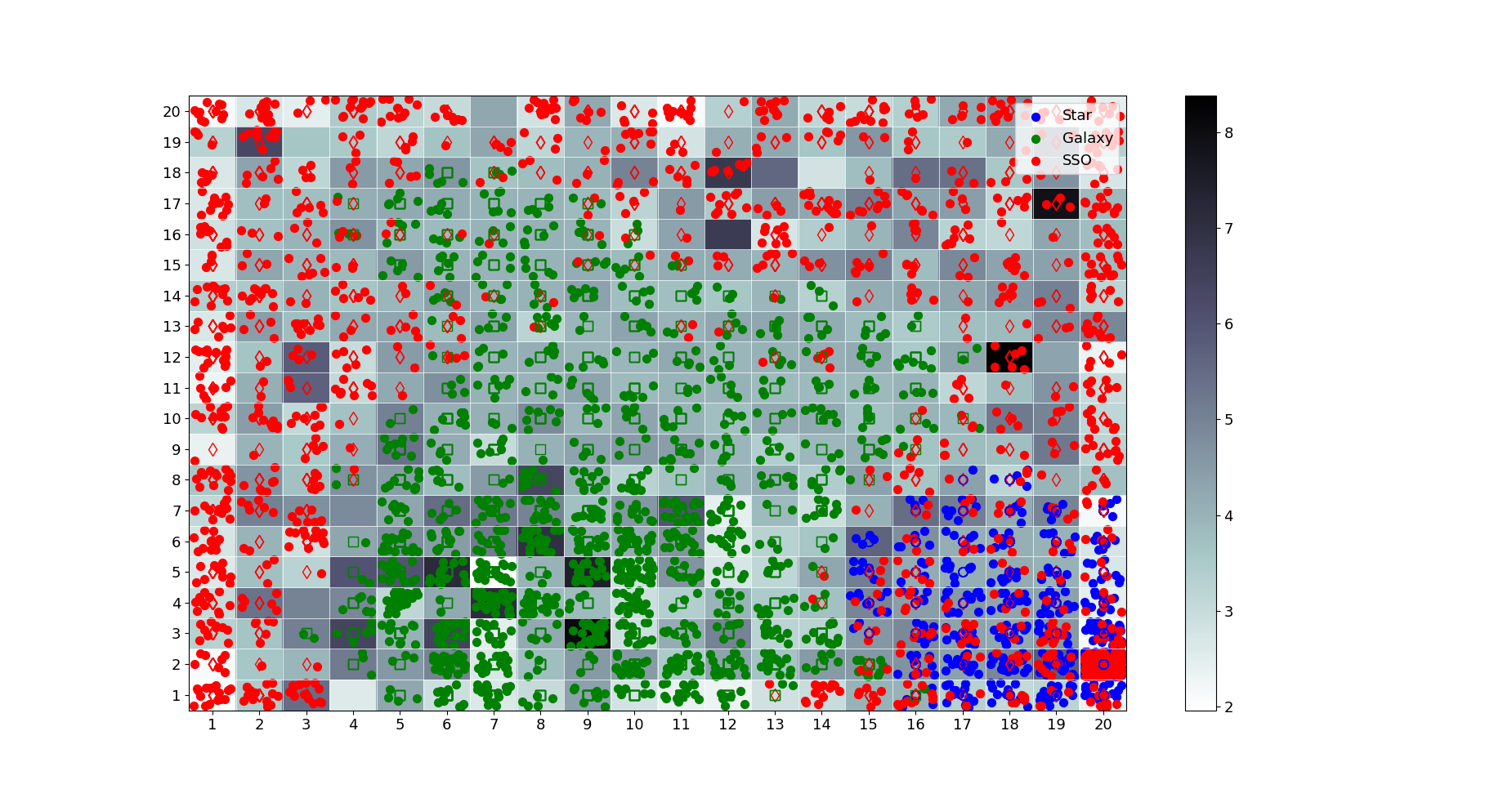}%
    \caption{20\,$\times$\,20 U-matrix of the SOM trained to classify stars, galaxies, and SSOs. Using the training data, we superimposed a different symbol and colour in the middle of each node for each type of input vector (blue circles for stars, green squares for galaxies, and red diamonds for SSOs), confirming that the SOM correctly classified the data. The dots appearing in each pixel (with a different colour for each class) indicate the number of objects of the particular class classified by the SOM. The underlying grey-scale image represents the U-matrix associated with the trained SOM, {while numbers along the axes represent the neuron position in the map.}
    %\BC{same comments: symbol and color-scale label}
    } 
    \label{result_2}%
  \end{figure*}
  \begin{figure*}
 \includegraphics[width=1.0\textwidth]{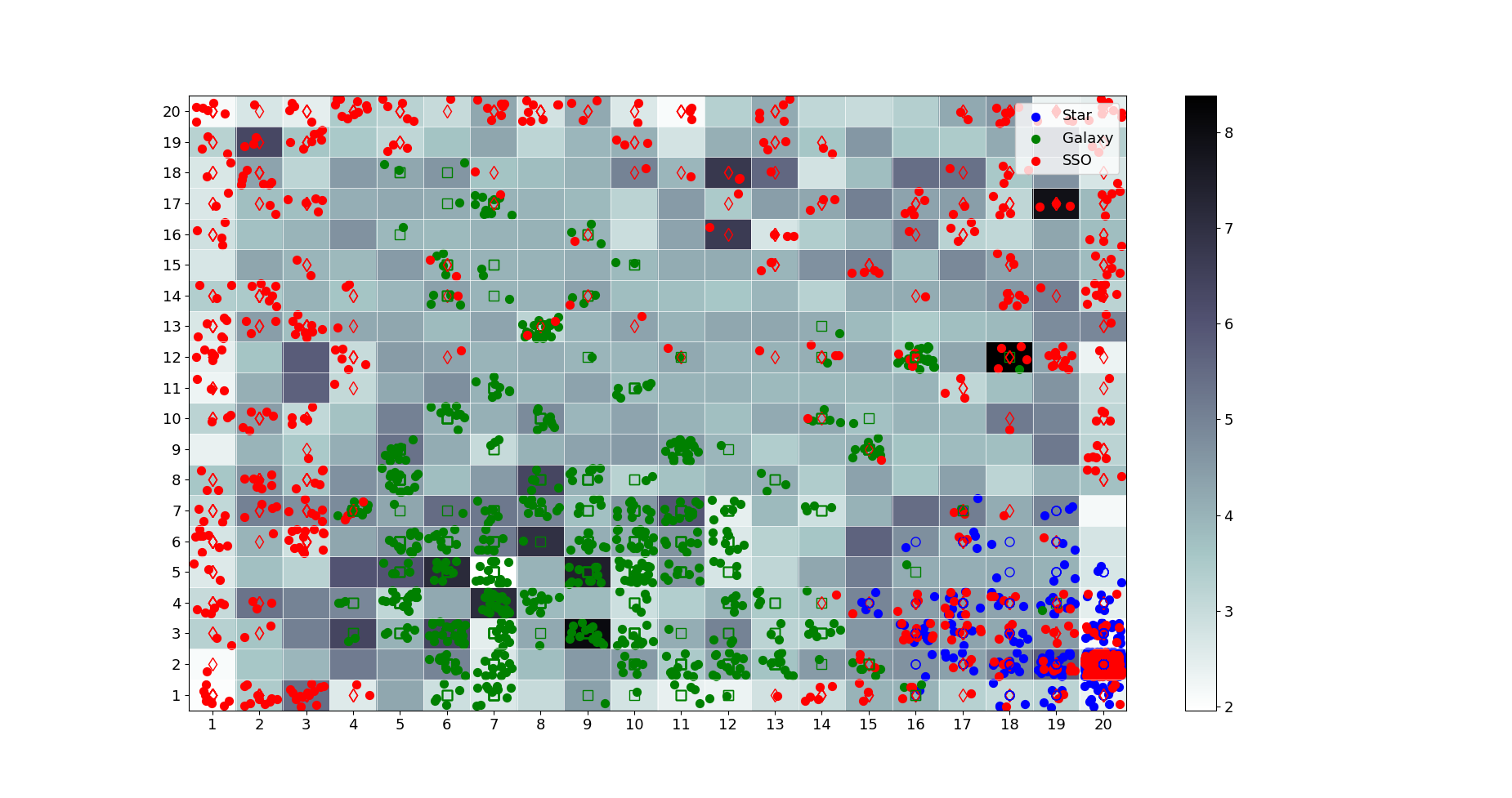}%
    \caption{20\,$\times$\,20 U-matrix of the same SOM given in Fig.~\ref{result_2} but exposed to a test data set. Having the correct class of the test data, we flagged the objects falling in each neuron with a different symbol and colour, and thus confirmed the capability of the (early stopped) SOM to classify a new data set. The meaning of the symbols and colours is as described in Fig.~\ref{result_2}.}
    \label{result_3}%
  \end{figure*}
  
  For any starting $\alpha$ parameter, the left panels show a monotonic improvement (decrease) in the error statistic, $S(t)$, which corresponds to a decrease in the $\sigma(t)$ value. During this phase, the SOM organises itself and finds clusters of similar data until it reaches a steady state at the end of the learning procedure. This is also reflected in the right panels where, for each selected pair of the $\alpha$ starting value and decreasing function, the $N_{\rm move}$ statistic is given. As was expected, the number of samples that change location in the map is large in the early stages of the training when the SOM is still creating large clusters, and then approaches zero, which indicates few movements during the fine-tuning of the weights or no movement at all, when the SOM has been trained. We note that $N_{\rm move}$ depends on the starting value of the learning parameter, $\alpha$. In particular, for $\alpha =0.1$ and $\alpha =0.5$, when adopting a linear decreasing function, the clusters continue to move around the map ($N_{\rm move}$ remains relatively large) until the SOM stabilises. Although the SOM converges on practically the same error, large movements at all epochs are unlikely in a real application. Therefore, we prefer the behaviour of the SOM trained by using the exponential decaying function (second and fourth rows in Fig.~\ref{result_1}). In this respect, the application of the early-stopping method implies that a general SOM becomes organised enough to classify new data after approximately $150$ epochs, and we prefer the SOM decaying as an exponential with $\alpha = 0.5$, because of its slightly smaller error reached when classifying the test data.

 \begin{figure}
     \hspace{-1.1cm} \includegraphics[width=0.6\textwidth]{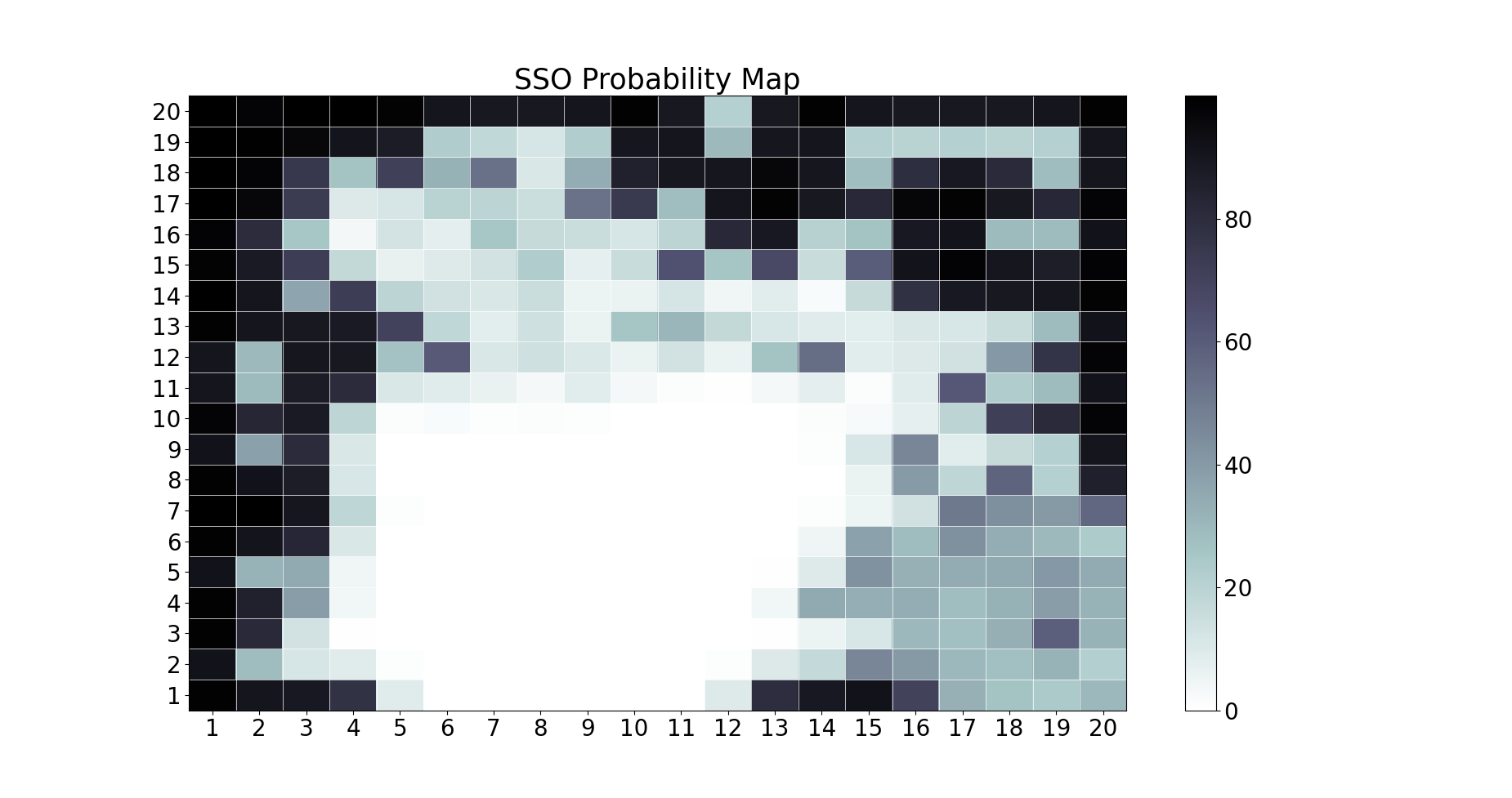}%
    \caption{SSO probability map (for a $20\times 20$ SOM) associated with the trained neural network (see text for details). The value of each pixel (according to the associated colour bar) gives the normalised probability that the data falling in that particular neuron belongs to the SSO class. The probability is normalised so that by summing up the probability per class, one gets exactly 100\%.}
    \label{result_4-a}%
  \end{figure}

 \begin{figure}
    \hspace{-1.1cm} \includegraphics[width=0.6\textwidth]{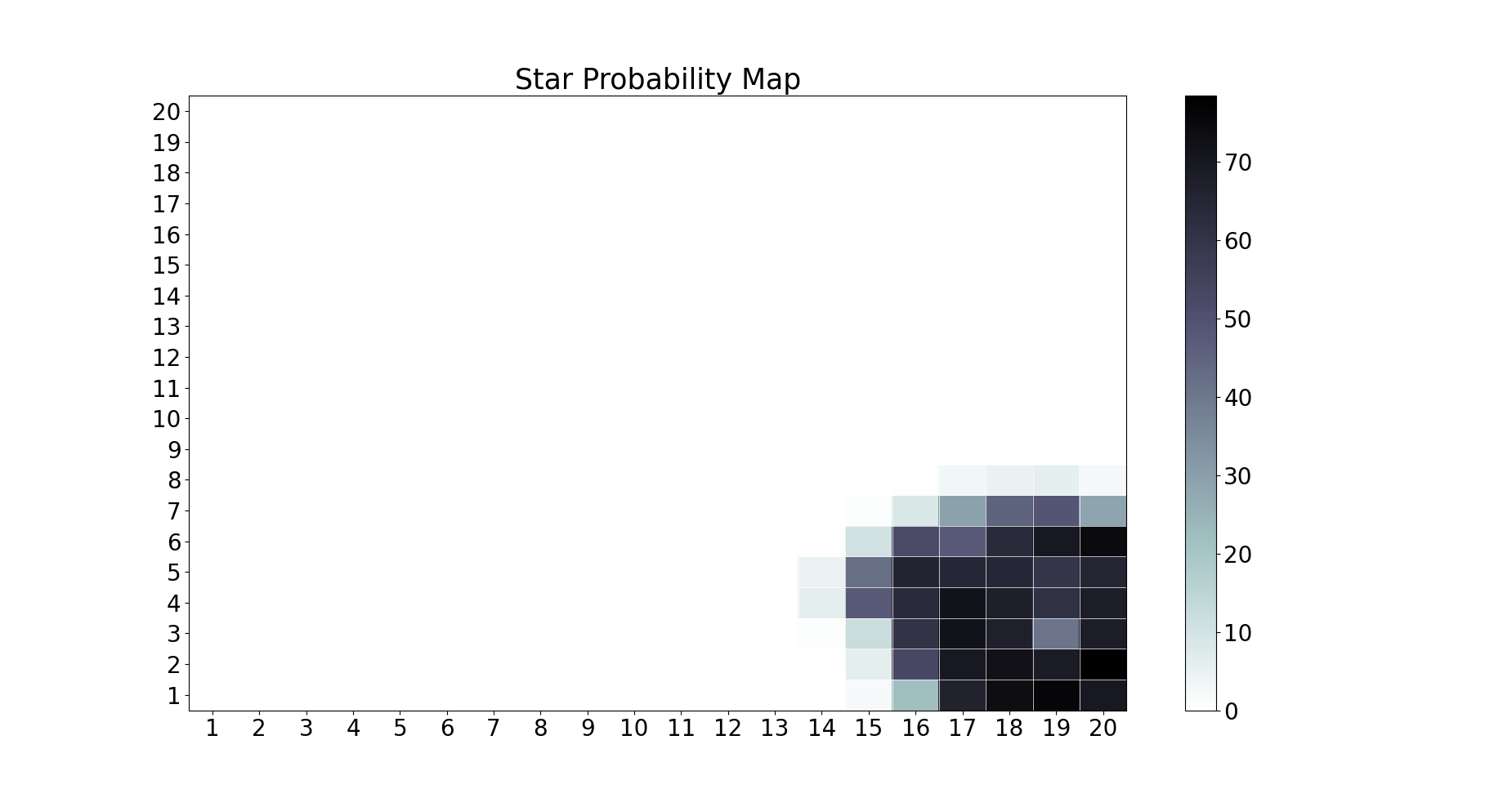}%
    \caption{Same as in Fig. \ref{result_4-a} but for the star class.}
    \label{result_4-b}%
  \end{figure}

 \begin{figure}
     \hspace{-1.1cm} \includegraphics[width=0.6\textwidth]{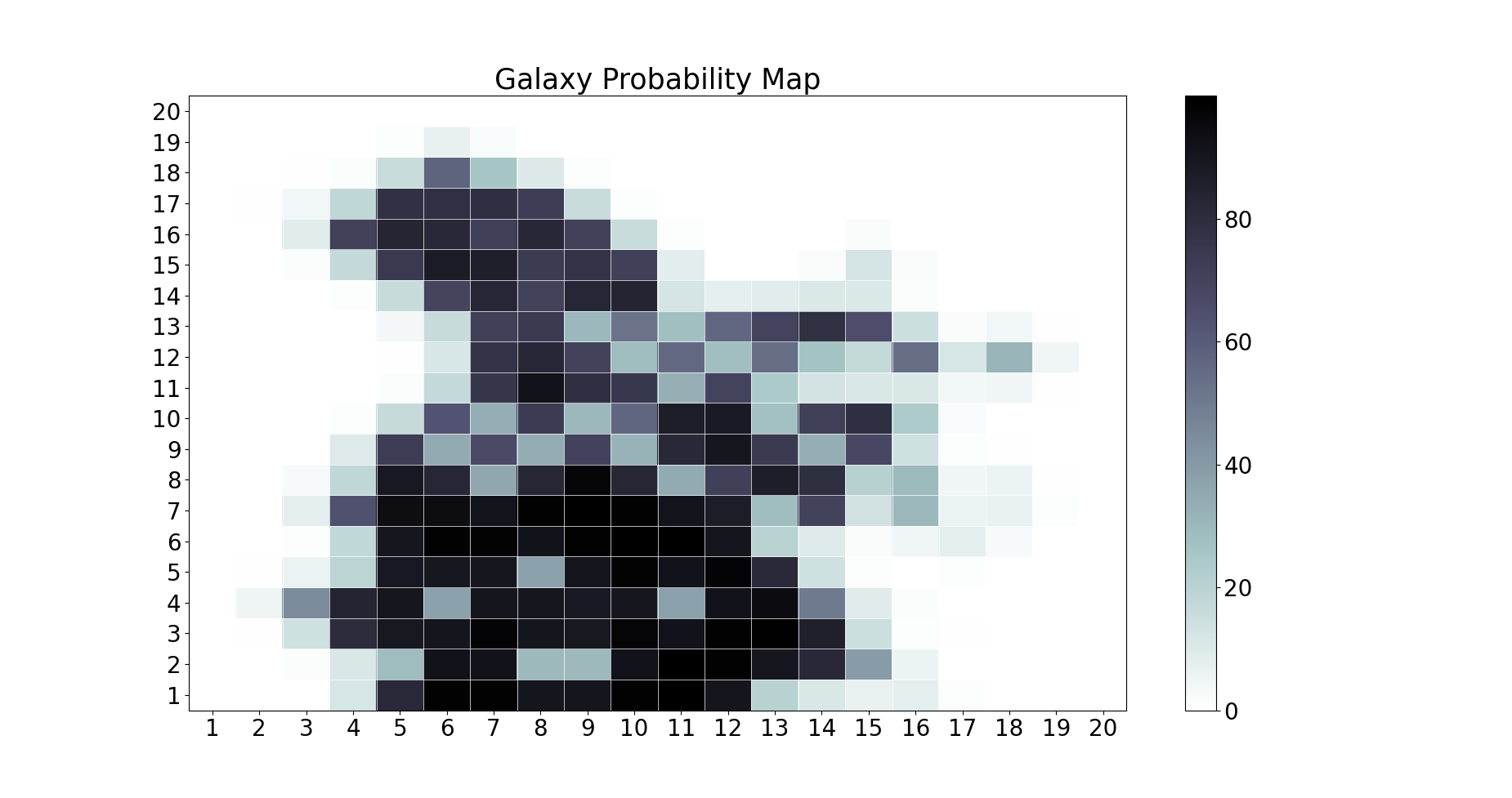}%
    \caption{Same as in Fig. \ref{result_4-a} but for the galaxy class.}
    \label{result_4-c}%
  \end{figure}
  
  After selecting the SOM architecture, we started training on the input data again with early stopping, as was discussed before, but this time allowing (for each run) a different initialisation of the neuron weights. We then selected the SOM with the smallest error in the epoch of over-fitting, $t_{\rm over-fitting}$.

  Finally, we built the associated U-matrix and projected onto each pixel the members of the input data set for which we had an a priori classification. Figure~\ref{result_2} shows the 20\,$\times$\,20 U-matrix superimposed with a different symbol and colour for each type of input vector (stars, galaxies, and SSOs), showing that the SOM succeeded in classifying the data correctly. The co-ordinates of the objects falling in the same neuron have been randomised within the same pixel for graphical purposes. In this respect, we note that the SSOs (red circles) that fall in the SOM region prevalently associated with stars (blue circles) are slow-speed objects that appear as point sources (characterised by a velocity below $1\arcsec\,{\rm h}^{-1}$) so that the SOM hardly separates them from the rest of classes. A further inspection of the same figure reveals that SSOs are spread over a large map area. This occurs since the neural network autonomously clusters the SSOs with similar lengths and orientations in the training images. Analogously, the SOM organises the galaxies, spreading them depending on the size and shape, placing those characterised by a large ellipticity at the outskirts of the long SSO region, as expected. In Fig.~\ref{result_3}, we give the results of the trained SOM exposed to a new test data set consisting of 2000 images with stars, galaxies, and SSOs for which we already have an a priori classification. As can be seen from the superimposed objects, the early-stopped SOM can adequately classify the new data. The underlying grey-scale image represents the U-matrix associated with the trained SOM.

\subsection{Self-organising-map probability}
\label{somprob-section}
  
  Since our final goal is to apply the trained SOM to a new data set (for which, of course, no a priori classification is given), we can associate with the neural network a probability of having an object type (picked up from a particular data set, $x$) in a given map neuron. This probability, $P_{i,j}^{x}$, is simply given by the ratio between the number of objects of a particular class falling in a neuron and the total number of test samples found in the same neuron, regardless of the associated type, as
  \begin{equation}
    P_{i,j}^{x, {\rm class}}= \frac{N_{i,j}^{\rm class}}{N_{i,j}}\;,
  \label{prob_1}
  \end{equation}
  where the class can be either star, galaxy, or SSO. 

  Although the general behaviour of the trained SOM is fixed, the probability, as is defined above, might be characterised by fluctuations depending on the data set passed to the neural network. Therefore, we simulated $X=30$ data sets (each of which contains $2000$ different objects), required that every class be represented by approximately 1/3 of the sample number, and gave the data sets to the SOM for a blind classification. For each data set, we then evaluated the probability per node for a given class, $P_{i,j}^{x, {\rm class}}$, and then averaged the results over the number, $X$, of data sets; in other words,
  \begin{equation}
    P_{i,j}^{{\rm class}}= \frac{\sum _{x=0}^{X-1} P_{i,j}^{x, {\rm class}}}{X}\;.
  \label{prob_2}
  \end{equation}
  
  {The results of the above calculation are reported in Figs.~\ref{result_4-a}, \ref{result_4-b}, and \ref{result_4-c}, where we give the probability that the SOM classifies input data into the SSO, star, and galaxy classes, respectively. The maps are normalised so that summing the class probabilities gives exactly 100\%.

  The usefulness of such maps is evident when trying to classify new input data with the trained SOM. In particular, when data is injected into the SOM and a winning node is identified, the probability per class associated with that neuron can be retrieved. We found that, although a trained SOM is capable of correctly classifying new images containing stars, SSOs, and galaxies with a certain degree of accuracy, SSOs characterised by a very slow speed (namely below $1\arcsec\,{\rm h}^{-1}$) are visually indistinguishable from point-like stars, and the SOM fails to classify them correctly. This is also clear from the classification map given in Fig.~\ref{result_3}, where star versus SSO confusion appears in the bottom right corner of the SOM.}

   The final recipe for determining the detection quality turns out to be:
  \begin{enumerate}[topsep=0pt]
    \item{For each source detected as a SSO by the \texttt{SSO-PIPE} pipeline (see Sect.~\ref{sec:pipe}), an image of 101\,$\times$\,101 pixels (centred on the target co-ordinates) is extracted. The image is then rebinned by a factor of three so that its cardinality is $K$.}
    \item{By using the trained SOM (with $N=M=20$ per side), a winning node (characterised by $k$ components) is identified for any input image.}
    \item {The probability maps in Figs.~\ref{result_4-a}, \ref{result_4-b}, and \ref{result_4-c} are then queried at the winning neuron co-ordinates, and the probability (per class) is extracted. This probability is then interpreted as the quality flag associated with the input target. Alternatively, a given input object is associated with the classification corresponding to the largest extracted probability.}
\end{enumerate}

{To assess the overall quality of the SOM, we followed the previous recipe and fed the SOM with a fresh validation set consisting of $10^4$ images (each including one amongst stars, galaxies, and SSOs simulated as is described in Sect.~\ref{sec:simu} and with a priori knowledge of the classification). We then evaluated the behaviour of the neural network by calculating the purity and completeness of the results (see Eqs. \ref{purity}--\ref{completeness}) as a function of the SSO speed but regardless of the object magnitude. For simplicity, we explicitly considered three bins of velocity (namely, $0\arcsec$--$3\arcsec\,{\rm h}^{-1}$,  $3\arcsec$--$6\arcsec\,{\rm h}^{-1}$, and $6\arcsec$--$10\arcsec\,{\rm h}^{-1}$) and found purity (completeness) values of $15\%$ ($70\%$),  $100\%$ ($60\%$), and $100\%$ ($58\%$) in the first, second, and third bins, respectively. This behaviour is expected, since SSOs with velocity values that fall in the first bin (and in particular those with velocity smaller than $0\arcsecf5\,{\rm h}^{-1}$) are not correctly recognised by the SOM but misclassified as stars, being formally indistinguishable from a fixed point-like source in a single image. 
}

\section{Results and discussion}
\label{sec:conclusions}

  In this paper, we describe the main features of the \texttt{SSO-PIPE}, a software developed and maintained at the \euc Science Operation Centre in ESAC/ESA, dedicated to the detection (and classification) of slow-moving SSOs with typical speeds lower than $10\arcsec\,{\rm h}^{-1}$ in \euc VIS images.
  
  %As described in Sect.~\ref{sec:pipe}, the detection algorithm is based on %the image astrometric registration (against a sample of Gaia stars) and &extraction of source catalogues. The output catalogues are then queried %for tracklets, i.e., sources that appear to move with respect to fixed %objects.

  \texttt{SSO-PIPE} shows high efficiency in recognising SSOs, particularly within the speed range of $0\arcsecf5 - 9\arcsec\,{\rm h}^{-1}$ and in magnitude bins where sources appear brighter, specifically up to $24$--$25$ in magnitude. Analysing the simulated observations with 2000 SSOs per field returns an approximate probability of 80\% of detecting a moving object. The probability that a detected object is a genuine SSO can be as large as $90\%$, indicating a high purity level. However, detecting fainter objects (falling in the last bin of investigated magnitude, $25$--$26$ ) fails in most cases. This is particularly true for faster objects characterised by very elongated streaks, which tend to blend into the background. Furthermore, as is seen in Figs.~\ref{fig:purity} and \ref{fig:completeness}, the detection and recognition of high-speed objects, specifically those approaching the $10\arcsec\,{\rm h}^{-1}$ limit, shows a decay due to inaccurate estimates of the target co-ordinates caused by the source fragmentation. This fact leads to a failure to detect objects whose velocity and position angle accuracy do not fall in the corresponding uncertainty limits considered in the pipeline. Conversely, detecting objects with speeds $<0\arcsecf5\,{\rm h}^{-1}$ is challenging, since very slow-moving objects appear almost stationary and are therefore confused with stars.

  {In this paper, we have also tested the possibility of using an SOM to separate the observed objects into star, galaxy, and SSO classes. Working on simulated data, we have shown that an SOM can be used as a classifier for new data sets if trained with the early-stopping technique. The SOM can recognise SSOs on realistic simulated images with relatively high accuracy, provided the object has a speed higher than $1\arcsec\,{\rm h}^{-1}$. Below this value, the neural network becomes practically useless in classifying the examined object, which appears formally indistinguishable from a fixed point-like source in a single image. 

  We have also verified that the SOM, if correctly trained, can disentangle between point-like stars and galaxies. In particular, as is implemented in this work, the trained SOM is capable of correctly classifying galaxies when characterised by a Sérsic index in the range of $0.5$--$4$. Contrarily, any increasing shape complexity introduces noise into the classification. As a matter of fact, there is always at least a difference between simulated and authentic images, and therefore, to work optimally with real \euc data, the SOM probably requires retraining with a new training set consisting partially or entirely of real training examples. Furthermore, the SOM is a particular neural network that recognises one pattern at a time. Therefore, under the assumption that the SOM is fed with an image containing multiple objects (for example, two sources belonging to different classes), the neural network will interpret the input as a particular noisy example of one class, with the non-retrieved object as a source of noise for the recognised one. 
  Investigating the source blending in the SOM training and forecasting remains work for the future. 
   
   Finally it is worth noting that this approach for the detection and classification of SSOs in true \euc VIS images (acquired during the week-long Phase Diversity Calibration campaign specifically planned towards the ecliptic for Solar System science) is currently in progress. This involves accurate fine tuning of all the detection parameters as well as retraining the presented neural network with real cases.
  }

\begin{acknowledgement}
  We are grateful for support from the ESAC (ESA) Science faculty and the INFN projects TAsP and EUCLID.

  This work is (partially) supported by ICSC – Centro Nazionale di Ricerca in High Performance Computing, Big Data and Quantum Computing, funded by European Union – NextGenerationEU.

  \AckECon
    
\end{acknowledgement}

%
% Here comes the reference list, generated via bibtex from
% the bibfile AandA.bib
%

\bibliography{biblio}

\label{LastPage}
\end{document}